\documentclass[reprint,twocolumn,superscriptaddress,showpacs,nofootinbib,notitlepage,preprintnumbers]{revtex4-2}

\usepackage{amsmath}
\usepackage{array}
\usepackage{graphicx}
\usepackage{hyperref}
\usepackage[dvipsnames]{xcolor}
\usepackage{amsmath,amsthm,amssymb}
\usepackage{comment}
\usepackage{enumerate}

\newcommand{\be}{\begin{equation}}
\newcommand{\ee}{\end{equation}}
\newcommand{\bea}{\begin{eqnarray}}
\newcommand{\eea}{\end{eqnarray}}

\begin{document}
\preprint{INT-PUB-24-015}

\title{Neutrino many-body flavor evolution: the full Hamiltonian}

\author{Vincenzo Cirigliano}
\affiliation{Institute for Nuclear Theory, University of Washington, Seattle, WA 98195, USA}
\author{Srimoyee Sen}
\affiliation{Department of Physics and Astronomy, Iowa State University, Ames, IA, 50011}
\author{Yukari Yamauchi}
\affiliation{Institute for Nuclear Theory, University of Washington, Seattle, WA 98195, USA}

\date{\today}

\begin{abstract}
We study neutrino flavor evolution in the quantum many-body approach  using the full neutrino-neutrino Hamiltonian,   including the usually neglected terms that mediate non-forward scattering processes.   
Working in the occupation number representation with plane waves as single-particle states, we explore the time evolution of simple initial states with up to $N=10$ neutrinos. 
We discuss the time evolution of the  Loschmidt echo, 
one body flavor and kinetic observables, 
and the one-body entanglement entropy.  For the small systems considered, 
we observe `thermalization' of both flavor and momentum degrees of freedom on comparable time scales, 
with results converging towards expectation values computed within a  microcanonical ensemble.
We also observe that the inclusion of  non-forward processes generates 
a faster flavor evolution compared to the one induced
 by the truncated (forward) Hamiltonian. 

\end{abstract}

\maketitle


\section{Introduction}

The evolution of neutrino flavor in hot and dense media provides key input to 
our understanding of  the synthesis of  light nuclei in the early universe 
and  heavy nuclei in the collapse or merger of  compact astrophysical objects, 
and affects the neutrino signal from a future galactic supernova. 
Astrophysical neutrinos are usually studied  through so-called Quantum Kinetic Equations (QKEs), 
which are evolution equations for  the one-body reduced neutrino density matrix, 
accounting for both momentum and flavor degrees of freedom~\cite{McKellar:1992ja,Sigl:1993ctk,Vlasenko:2013fja,Volpe:2013uxl,Froustey:2020mcq}. 
The QKEs have been derived  from quantum field theory using various methods, 
including the two-particle-irreducible (2PI) effective action~\cite{Berges:2015kfa} 
truncated to three loops~\cite{Vlasenko:2013fja,Blaschke:2016xxt}.
The QKEs involve both coherent forward scattering and collisional kernel,  
and lead to a rich phenomenology of collective phenomena and flavor instabilities 
(see the review \cite{Volpe:2023met} and references therein).

While the computational implementation of the full QKEs and their interface with compact objects evolution codes   
is an arduous task,   understanding the limits of applicability of the one-body approach underlying 
the QKEs remains an active area of research.  The question whether the one-body analysis leaves out 
important many-body correlations and entanglement effects goes back to the early days of 
the field~\cite{Pantaleone:1992xh,Pantaleone:1992eq} and has received attention over the years~\cite{Bell:2003mg,Friedland:2003dv,Friedland:2003eh,Sawyer:2005jk,
Friedland:2006ke,McKellar:2009py}.  

More recently, the  validity of the QKE treatment of the neutrino gas has  
come under scrutiny  in the context of quantum many-body  approaches to this problem 
(see \cite{Balantekin:2023qvm}  and references therein). 
As far as we are aware, all existing quantum many-body studies of the neutrino system 
use a truncated Hamiltonian  $H_{\nu \nu}^{(T)}$~\cite{Pehlivan:2011hp} 
that only couples pairs of  momentum-space neutrino operators satisfying  forward kinematics.
In other words, $H_{\nu \nu}^{(T)}$ contains  terms  
that either preserve or exchange the momenta of 
interacting neutrino pairs (forward and exchange terms). 
As argued in  Ref.~\cite{Johns:2023ewj},  the use of the truncated Hamiltonian 
is not justified in a first-principles many-body approach.
On the other hand,  the truncated $\nu$-$\nu$ Hamiltonian has the virtue of 
mapping onto a  spin-spin  Hamiltonian with all-to-all couplings~\cite{Pehlivan:2011hp}, 
which is amenable to many-body analyses~\cite{Birol:2018qhx,Rrapaj:2019pxz,Cervia:2019res,Roggero:2021asb,Roggero:2021fyo,Xiong:2021evk,Martin:2021bri,Roggero:2022hpy,Lacroix:2022krq,Martin:2023ljq,Martin:2023gbo,Bhaskar:2023sta} and  
implementation on quantum computers~\cite{Hall:2021rbv,Amitrano:2022yyn,Illa:2022zgu,Nguyen:2022snr,Siwach:2023wzy}.
In certain regimes, the many-body results are at variance with the QKE expectations (see for example Refs.~\cite{Balantekin:2023qvm,Martin:2023ljq,Martin:2023gbo}), 
and there is an ongoing debate on whether these calculations 
can indeed challenge the validity of the QKEs approach~\cite{Shalgar:2023ooi,Johns:2023ewj,Kost:2024esc}. 

The path towards more realistic many-body studies requires 
several  developments, which include: 
(i) assessing the impact of  using the full many-body Hamiltonian rather than its truncated version;  
(ii) exploring more general initial pure states  (not product states of plane waves), 
and possibly admixtures of pure states, that more realistically  describe  the physical system; 
(iii) studying the dynamics of larger systems and systematically studying the scaling 
of relevant observables with the number of neutrinos; 
including neutrino interactions with other particles.
In this work,  we address  point (i) above:  
first, we work out the full neutrino-neutrino Hamiltonian $H_{\nu \nu}$ 
and  set up the framework to implement the time evolution 
using the occupation number representation. 
Then we explore the evolution of  simple initial states. 
In this simplified setting, we study the time scales for 
evolution of flavor and momentum degrees of freedom, 
and their interplay. 
While we use a plane wave single-particle basis, we emphasize that 
in principle any initial state can be built within 
this formalism, thus allowing one to study point (ii) above. 

This paper is organized as follows: 
in Sect.~\ref{sect:formalism} we setup the basic formalism, 
present the full Hamiltonian and give its matrix elements 
in the occupation number basis. 
In Sect.~\ref{sec:rhoands} we introduce the one-body density matrix and 
the corresponding entanglement entropy, emphasizing the interplay 
between flavor and momentum degrees of freedom. 
In Sect.~\ref{sect:setup} we discuss the relevant energy scales in the 
problems of astrophysical interest. 
In Sect.~\ref{sect:toy} we study the time evolution in a toy model 
with $N=2$, illustrating some features that generalize to larger systems. 
In Sect.~\ref{sect:grid} we study the time evolution of neutrino systems with 
$N=6,8,10$ in a two-dimensional setup and investigate qualitative 
and quantitative differences that emerge when using the full and truncated Hamiltonian. 
We summarize our results in  Sect.~\ref{sect:discussion}  
and provide some more technical details in  appendices \ref{sect:appendix2} and  \ref{sect:app_nubar}.


\section{Formalism}
\label{sect:formalism}

In order to write down the Hamiltonian 
we describe neutrino fields $\nu_\alpha(x)$ as four-component spinors  
($\alpha \in \{e, \mu, \tau \}$ denoting the flavor). At energies much smaller than the electroweak scale the Hamiltonian takes the form  
(repeated flavor indices are summed over) 
\begin{eqnarray}
\label{eq:Hf}
H &=& H_{\rm kin} + H_{\nu \nu}  + H_{\nu-{m}}
\end{eqnarray}
with
\begin{eqnarray}
\label{eq:H}
H_{\rm kin} &=&   \int d^3x \ \bar{\nu}_\alpha (x) \left( -i \delta_{\alpha \beta} {\bf \gamma} \cdot {\bf \nabla}  + m_{\alpha \beta} \right) \nu_\beta (x), \,\,\,\,\,\,\nonumber\\
H_{\nu \nu}
&=&  \frac{G_F}{\sqrt{2}} \int d^3x  \  \bar{\nu}_\alpha (x) \gamma_\mu P_L  \nu_\alpha (x) \  \,  
\bar{\nu}_\beta (x) \gamma^\mu P_L  \nu_\beta (x) ,  \quad
\end{eqnarray}
where we will use the following gamma matrices
\bea
\gamma^0=\begin{pmatrix}
0 && \sigma^0\\
\sigma^0 && 0
\end{pmatrix}&,&
\gamma^i=\begin{pmatrix}
0 && \sigma^i\\
-\sigma^i && 0
\end{pmatrix}\nonumber\\
\gamma_5&=&i\gamma^0\gamma^1\gamma^2\gamma^3
\eea
with $\sigma^0$ being the two dimensional identity matrix and $\sigma^i$ the $i^{\text{th}}$ Pauli matrix. $P_L = (1 - \gamma_5)/2$ and 
$m$ is a complex mass matrix for Dirac neutrinos (for Majorana neutrino the  
 kinetic term acquires an overall  factor of $1/2$ and the mass matrix becomes symmetric).  
$H_{\nu-m}$ denotes the interaction of neutrinos with quarks and charged leptons. For a complete description of neutrino-matter interaction see 
for example \cite{Blaschke:2016xxt}. In this work we do not consider the effects of $H_{\nu-m}$. 
In what follows we expand the neutrino fields in creation and annihilation  operators 
and derive a representation of the Hamiltonian in Fock space. 


\subsection{Neutrino fields and spinors}
\label{sect:appendix}

We expand the free Dirac neutrino fields as follows 
\begin{eqnarray}
\nu_{i} (x)
&=&  \sum_{h=\pm} \  \int \frac{d^3 {\bf p}}{(2 \pi)^3}  \, 
  \bigg( u ({{\bf p},h})  a_i ({{\bf p},h})  e^{-ipx} 
 \nonumber \\ 
  & & \ \ \ \ \ + \ v ({{\bf p},h})  b_i^\dagger ({{\bf p},h}) e^{ipx} \bigg)
\end{eqnarray}
in terms of helicity spinors $u ({{\bf p},h})$,  $v ({{\bf p},h}) $ and 
creation  / annihilation operators for neutrinos  ($a_i ({{\bf p},h})$) and antineutrinos ($b_i ({{\bf p},h})$). 
The $h \in \{+,-\}$ label refers to helicity and $i \in \{1,2\}$ refers to the mass eigenstate. 
With the normalizations adopted here, the creation and annihilation operators carry mass dimension ${-3/2}$ and 
 satisfy the following anti-commutation relations: 
\begin{equation}
\label{eq:canonical}
\{ a_\alpha ({\bf p}, h) , a^\dagger_\beta ({\bf p}^\prime, h^\prime) \} =   (2 \pi)^3   \delta^{(3)} ( {\bf p} - {\bf p}^\prime)   \, \delta _{h h^\prime} \, \delta_{\alpha \beta}~. 
\end{equation}

In our conventions the spinors are dimensionless and normalized such that 
\begin{equation}
u^\dagger  ({{\bf p}, h}) u  ({{\bf p}, h^\prime})  = v^\dagger  ({{\bf p}, h})  v ({{\bf p}, h^\prime})  = \delta_{h h^\prime}~. 
\end{equation} 
The helicity 4-spinors are given by:
\be
u({\bf p},+) =  \sqrt{\frac{E + |{\bf p}|}{2E}} \  \left( 
\begin{array}{c}
r({\bf p})  \, \xi_+ (\hat{\bf p}) \\
\xi_+ (\hat{\bf p})
\end{array}
\right) ~, 
\ee
\be
u({\bf p},-) =  \sqrt{\frac{E + |{\bf p}|}{2E}} \ 
 \left( 
\begin{array}{c}
 \xi_- (\hat{\bf p}) \\
r ({\bf p}) \,  \xi_- (\hat{\bf p})
\end{array}
\right)~,
\ee
\be
v({\bf p},+) =  \sqrt{\frac{E + |{\bf p}|}{2E}} \  \left( 
\begin{array}{c}
\xi_- (\hat{\bf p}) \\
- r({\bf p}) \, \xi_- (\hat{\bf p})
\end{array}
\right) ~, 
\ee
\be
v({\bf p},-) =  \sqrt{\frac{E + |{\bf p}|}{2E}} \ 
 \left( 
\begin{array}{c}
- r({\bf p}) \,  \xi_+ (\hat{\bf p}) \\
  \xi_+ (\hat{\bf p})
\end{array}
\right)~,
\ee
with 
$E = \sqrt{{\bf p}^2 + m^2}$ and  
$r({\bf p}) = {m}/(E +|{\bf p}|) = \sqrt{(E-|{\bf p}|)/(E+ |{\bf p}|)}$. 
Denoting by 
$\theta_{\bf p}, \phi_{\bf p}$ the polar and azimuthal angles of $\hat{\bf p} \equiv {\bf p}/|{\bf p}|$) the helicity 
Pauli spinors are 
\be
\xi_+ (\hat{\bf p}) = 
 \left( 
\begin{array}{c}
 \cos \frac{\theta_{\bf p}}{2} \\
e^{i \phi_{\bf p}}  \ \sin \frac{\theta_{\bf p}}{2} 
\end{array}
\right)~
\quad 
\xi_- (\hat{\bf p}) = 
 \left( 
\begin{array}{c}
- e^{- i \phi_{\bf p}}  \ \sin \frac{\theta_{\bf p}}{2} \\
 \cos \frac{\theta_{\bf p}}{2} 
\end{array}
\right)
\ee
and satisfy $( \vec{\sigma} \cdot \hat{\bf p} )  \ \xi_\pm (\hat{\bf p})  \ = \  \pm  \  \xi_\pm (\hat{\bf p})$. 

Throughout, we treat neutrino masses as perturbations and neglect terms of higher order in $m_i / |{\bf p}|$. 
As is well known, in this approximation the L-handed neutrino field appearing in the weak Hamiltonian only involves 
left-helicity neutrinos and right-helicity antineutrinos
\begin{eqnarray}
P_L \nu_{i} (x)
&=& \int \frac{d^3 {\bf p}}{(2 \pi)^3}  \, 
  \bigg( u ({{\bf p},-})  a_i ({{\bf p},-})  e^{-ipx} 
 \nonumber \\ 
  & & \ \ \ \ \ + \ v ({{\bf p},+})  b_i^\dagger ({{\bf p},+}) e^{ipx} \bigg).
\end{eqnarray}
Further  focusing on the many-body dynamics of neutrinos (ignoring antineutrinos for simplicity) 
we only need to consider terms involving left-helicity neutrino mode operators. 
To simplify the notation, we thus suppress the redundant helicity label: 
$a_i ({\bf p},-) \to a_i ({\bf p})$.

Finally, since the interaction Hamiltonian is most naturally expressed in terms of 
flavor fields, we  introduce the flavor basis mode operators 
and express the Hamiltonian in terms of these. 
In the two-flavor case, the relation between mass and flavor operators is given by 
\bea
a_e ({\bf p}) &=& \cos \theta \ a_1 ({\bf p})  + \sin \theta  \ a_2 ({\bf p}) 
\nonumber \\
a_\mu ({\bf p}) &=& - \sin \theta \ a_1 ({\bf p})  + \cos \theta  \ a_2 ({\bf p}) ~.
\eea


\subsection{Hamiltonian}

In what follows we focus on the many-body dynamics of neutrinos, ignoring antineutrinos for simplicity. 
For completeness, the terms in $H_{\nu \nu}$ involving anti-neutrinos are reported in Appendix~\ref{sect:app_nubar}.
As discussed above, we quantize the fields in the mass basis and 
work in the ultra-relativistic limit $m_i / |{\bf p}| \ll 1$. We then express 
the Hamiltonian  in terms of  creation and annihilation operators of left-helicity neutrinos,  
for which we use the flavor basis: 
$a_\alpha ({\bf p})$, $\alpha \in \{ e, \mu , \tau\}$. 
From now on, we restrict our discussion to the case of two flavors $e$ and $\mu$. The generalization 
to three flavors is straightforward. 

Defining  $\delta m^2 = m_2^2 - m_1^2$, the kinetic term of the Hamiltonian reads
\begin{align}
    H_{\rm kin} = & \int  \frac{d {\bf p}}{(2 \pi)^3} \nonumber\\
    &\bigg[  \left( |{\bf p}| +  \frac{m_1^2 + m_2^2 - \cos 2 \theta \, \delta m^2}{4 |{\bf p}|}  \right)  
 a_e^\dagger ({\bf p}) a_e ({\bf p})  
\nonumber \\
&+ 
 \left( |{\bf p}| +  \frac{m_1^2 + m_2^2 + \cos 2 \theta \, \delta m^2}{4 |{\bf p}|}  \right)  
 a_\mu^\dagger ({\bf p}) a_\mu ({\bf p})  
\nonumber\\
&+ \frac{\sin 2 \theta \ \delta m^2}{4 |{\bf p}|} 
\Big(  a_e^\dagger ({\bf p}) a_\mu  ({\bf p}) +  a_\mu^\dagger ({\bf p}) a_e  ({\bf p})  \Big)
\bigg]~,
\label{kin}
\end{align}

with the last term representing the usual vacuum mixing term. 

The  neutrino-neutrino terms in the interaction Hamiltonian $H_{\nu \nu}$  take  the form~
\begin{widetext}
\begin{eqnarray}
    H_{\nu \nu} &=  & \frac{G_F}{\sqrt{2}} \ 
    \sum_{\alpha, \alpha^\prime, \beta, \beta^\prime} \ 
         \int  \frac{d{\bf q}}{(2 \pi)^3}  \frac{d{\bf q^\prime}}{(2 \pi)^3}  \frac{d{\bf p} }{(2 \pi)^3} \frac{ d{\bf p^\prime}}{(2 \pi)^3} 
\      (2 \pi)^3 \delta ( {\bf p} + {\bf q}     - {\bf p^\prime} - {\bf q^\prime} ) 
\nonumber    \\ 
    & \times &  \  \Big(  a^\dagger_{\alpha^\prime} ({\bf p^\prime}) \, a_\alpha ({\bf p}) \, 
     a^\dagger_{\beta^\prime} ({\bf q^\prime}) \, a_\beta ({\bf q})    
\ \frac{(\delta_{\alpha^\prime \alpha} \delta_{\beta^\prime \beta} + \delta_{\alpha^\prime \beta} \delta_{\beta^\prime \alpha})}{2}
\ g ({\bf p^\prime}, {\bf p}, {\bf q^\prime}, {\bf q} ) 
      \  +  \  ...  \ \Big) ~, 
\label{eq:H2}
\end{eqnarray}
with 
\begin{align} 
 g ({\bf p^\prime}, {\bf p}, {\bf q^\prime}, {\bf q} )   
 & \equiv    \bar{u} ({\bf p}^\prime,-) \gamma_\mu P_L  u ({\bf p},-)  \
     \bar{u} ({\bf q}^\prime,-) \gamma^\mu P_L  u ({\bf q},-)  =  f^{\dagger}({\bf p}^\prime, {\bf q}^\prime)  ~f({\bf p}, {\bf q})     
     \label{eq:g} \\
f({\bf p}, {\bf q}) & = \sqrt{2} \, 
 \bigg( e^{- i \phi_{\bf p}}
\sin \left(\frac{\theta_{\bf p}}{2}\right) 
\cos \left(\frac{\theta_{\bf q}}{2}\right) 
- 
e^{-i \phi_{\bf q}}
\cos \left(\frac{\theta_{\bf p}}{2}\right) 
\sin \left(\frac{\theta_{\bf q}}{2}\right) 
\bigg)  ~.  \label{eq:f}
\end{align}
\end{widetext}

From Eq.~(\ref{eq:H2}) we see that $H_{\nu \nu}$ takes each pair of occupied states with momenta 
${\bf p}$, ${\bf q}$ to states with momenta 
${\bf p^\prime}$, ${\bf q^\prime}$, subject to the condition ${\bf p} + {\bf q} = {\bf p^\prime} + {\bf q^\prime}$,  
and acts on the flavors by either leaving them unchanged or by swapping them. 
The weight for each set of momenta considered is given by the function 
$ g ({\bf p^\prime}, {\bf p}, {\bf q^\prime}, {\bf q} ) $ in Eq.~(\ref{eq:g}).

Note that $g ({\bf p^\prime}, {\bf p}, {\bf q^\prime}, {\bf q} )= g ( {\bf q^\prime}, {\bf q}, {\bf p^\prime}, {\bf p} ) = - g ({\bf p^\prime}, {\bf q}, {\bf q^\prime}, {\bf p} )$,  where the last equality follows from the Fierz identities. 
In the forward scattering kinematics (${\bf p}^\prime = {\bf p}$ or ${\bf p}^\prime = {\bf q}$),  
this expression reproduces the familiar factors encountered in the 
literature~\cite{Pantaleone:1992eq}
\begin{align}
 g ({\bf p}, {\bf p}, {\bf q}, {\bf q} )    & =  - g ({\bf q}, {\bf p}, {\bf p}, {\bf q} )
 = 1 -  {\bf \hat p} \cdot {\bf \hat q}~.
 \label{eq:gforward}
\end{align}
As an illustration, we  consider the 2-dimensional case with ${\bf p}_y = 0$ and ${\bf p}_x > 0$ (corresponding to $\phi_{\bf p}=0$, $\phi_{\bf q}=0$), 
which leads to 
\be
 g ({\bf p^\prime}, {\bf p}, {\bf q^\prime}, {\bf q} )   \to 2 \sin \left(\frac{\theta_{\bf p} - \theta_{\bf q}}{2} \right) 
  \sin \left(\frac{\theta_{\bf p^\prime} - \theta_{\bf q^\prime}}{2} \right) ~.
  \label{eq:planar}
\ee
The one-dimensional case (all momenta along the $z$ axis) 
corresponds to $\theta_{{\bf p}, {\bf p^\prime}, {\bf q}, {\bf q}^\prime } = 0, \pi$, 
depending on the sign of the $z$-component of the momenta. 
Eq.~(\ref{eq:planar}) shows that in this case a non-zero amplitude is only obtained 
when the initial and final state momenta are `head on' (i.e. when the momenta 
satisfy momentum conservation ${\bf p}_z + {\bf q}_z = {\bf p^\prime}_z + {\bf q^\prime}_z$  and both 
conditions $sign ({\bf q}_z) = - sign ({\bf p}_z)$ and 
$sign ({\bf q^\prime}_z) = - sign ({\bf p^\prime}_z)$ hold). 

When working in finite volume, the formulae presented above need to be modified in the usual way. 
Assuming the box has linear size $L$ and volume $V = L^3$,  the 3-momenta ${\bf p}$ are uniquely identified 
by triplets of integers  
$({\bf z}_{\bf p})_{x,y,z}$
through  
$({\bf p})_{x,y,z}  =  [ (2 \pi)/L ] ({\bf z}_{\bf p})_{x,y,z}$.
As a consequence, the integrals over 3-momenta are replaced by finite sums over triplets of integers through the usual relation 
\be
\int \frac{d^3p}{(2 \pi)^3}  \to  \frac{1}{V} \sum_{{\bf z}_{\bf p}}~, 
\ee
and the Dirac delta functions of momentum conservation become Kroeneker deltas according to
\be  (2 \pi)^3  \delta^{(3)} ({\bf p} + {\bf q} - {\bf p}^\prime - {\bf q}^\prime) 
\to  V \  \delta_{{\bf z}_p+{\bf z}_q -{\bf z}_{p^\prime} - {\bf z}_{q^\prime,} {\bf 0}}.
\ee


\subsection{Fock space}
\label{sect:Fock}

We consider for simplicity only two neutrino flavors (denoted by $e$ and $\mu$) and  work in  Fock space. 
A single-particle state is identified by the three-momentum ${\bf p}_i$ ($i \in \{1,....,k\}$) 
and a flavor label $\alpha$ ($\alpha \in \{e,\mu\}$). 
We consider only neutrinos with negative helicity, so we don't have to specify any other quantum numbers. 
There are $2k$ single particle states and 
a basis vector  in Fock space is  specified by the set of occupation numbers 
$n_{i \alpha} \in \{0,1 \}$.  
The dimension of this space is $2^{2k}$~\footnote{When considering $n_f$ flavors, 
one simply makes the replacement $2k \to n_f \times k$.}.

We set out to study the problem in which the initial state 
has total number of neutrinos $N <k$:
\begin{equation}
    N = \sum_{\scriptsize \begin{array}{c} i=1,..,k \\ \alpha = e, \mu \end{array}}    
    n_{i \alpha}~. 
\label{eq:N}
\end{equation}
Since  $H_{\nu \nu}$ conserves the total number of neutrinos, we need to evolve 
the state in the space of fixed  $N$, which has  dimension  
\begin{equation}
    d_{N,k} = \left( 
    \begin{array}{c}
    2 k \\
    N
    \end{array}
    \right)~.
\end{equation}
The $d_{N,k}$ basis vectors  are labeled 
by 
\begin{equation}
{\bf n} = \{n_{1e}, n_{1\mu}, ... , n_{ke}, n_{k\mu} \}~,
\label{eq:narray}
\end{equation}
the 2$k$-dimensional array of occupation numbers 
obeying the condition (\ref{eq:N}), and represent anti-symmetrized products of $N$ single-particle states:
\begin{equation}
    | {\bf n} \rangle =  {\rm S.D.} \left( \prod_{i \alpha: \  n_{i\alpha} = 1}  \, | {\bf p}_i, \alpha   \rangle \right)~,
\label{eq:basis}
\end{equation}
where ``S.D." stands for Slater Determinant. 
A generic state is specified by  $d_{N,k}$ complex amplitudes $c_{{\bf n}}$ as follows
\begin{equation}
|\Psi \rangle = 
    \sum_{{\bf n}} 
    \ c_{{\bf n}} \, |  {\bf n} \rangle~.
    \label{eq:MBstate}
\end{equation}

Finally, to take into account the anti-commutation of the creation and annihilation operators correctly, 
we need to introduce the ordering rule for the operators in defining the basis vectors $|  {\bf n} \rangle$ in Eq.~(\ref{eq:basis}).   
The basis vectors $|  {\bf n}\rangle$ are defined via the application of a sequence of creation operators ordered 
in an increasing order of flavor and momenta, 
i.e. momenta with a smaller label are on the left, and  within a given momentum label 
the electron flavor goes to the left of the muon flavor. 
For example, the  
normalized 
basis state of three neutrinos with flavor and momenta $({\bf p}_1, e),({\bf p}_1, \mu),({\bf p}_2, e)$ 
(labeled by  ${\bf n}$ with $n_{1 \mu} = n_{2e}= n_{1\mu} = 1$ and all other occupation 
numbers $n_{j \beta} = 0$) is given by 
\begin{equation}
    |{\bf n}\rangle = \frac{a^{\dagger}_{e}({\bf {p}}_1)}{\sqrt{V}}~\frac{a^{\dagger}_{\mu}({\bf {p}}_1)}{\sqrt{V}}~\frac{a^{\dagger}_{e}({\bf {p}}_2)}{\sqrt{V}}~ |0\rangle \;\text.
\end{equation}
Its complex conjugate is defined as 
\begin{equation}
     \langle {\bf n} |    
    = \langle0|  ~ \frac{a_{e}({\bf {p}}_2)}{\sqrt{V}}~\frac{a_{\mu}({\bf {p}}_1)}{\sqrt{V}}~\frac{a_{e}({\bf {p}}_1)}{\sqrt{V}}\;\text.
\end{equation}
This defines an ortho-normal basis, i.e. $\langle {\bf n}| {\bf m} \rangle = \delta_{{\bf n},{\bf m}}$.
The application of an annihilation operator $a_{\alpha}({\bf  p}_i) $ to a basis vector $|{\bf n}\rangle$ results in
\begin{equation}
    a_{\alpha}({\bf {p}}_i) |{\bf n}\rangle = V^{1/2} f_{{\bf n},i,\alpha} ~\delta_{n_{i,\alpha},1} ~|{\bf n}^{[i \alpha]}\rangle
\label{eq:ann1} 
\end{equation}
where 
\begin{equation}
    {\bf n}^{[i\alpha]} = {\bf n} ~~\mathrm{with}~~ n_{i,\alpha}\rightarrow 0
\end{equation}
and 
\begin{equation}
    f_{{\bf n},i,\alpha} = (-1)^{\sum_{(j,\beta)<(i,\alpha)} n_{j,\beta}}\;\text. 
\end{equation}
The volume factor in Eq.~(\ref{eq:ann1}) arises due to the normalization adopted for the creation and annihilation operators 
(see Eq.~(\ref{eq:canonical}) and its finite volume version). 
The summation $\sum_{(j,\beta)<(i,\alpha)} n_{j,\beta}$ in the exponent of the anti-commutation factor $f_{\{n\},i,\alpha}$ means that we sum $n_{j,\beta}$ for all $(j,\beta)$ that are on the left of $(i,\alpha)$ in the ordering rule introduced above. 


\subsection{Matrix elements of the Hamiltonian}

We next discuss in some detail  the matrix elements of the Hamiltonian 
$H_{\mathrm{Kin}}$ and $H_{\nu \nu}$ in the occupation number basis.  
With the notation introduced in the previous subsection,
the matrix element of quadratic operators of the form $a^{\dagger}_{\alpha}({\bf  p}_i)a_{\beta}({\bf  p}_j)$ is
\begin{align}\label{eq:quad_me}
 \frac{1}{V}  & \ \langle {\bf  m} | ~a^{\dagger}_{\alpha}({\bf {p}}_i)~a_{\beta}({\bf {p}}_j)~| {\bf  n} \rangle 
\nonumber \\
=&  
\ \Bigl(f_{{\bf m},i,\alpha} ~\delta_{m_{i,\alpha},1} \Bigr)
    \Bigl( f_{{\bf n},j,\beta} ~\delta_{n_{j,\beta},1}\Bigr) \langle { \bf m }^{[i\alpha]} | {\bf n}^{[j\beta]} \rangle  
\nonumber \\
\equiv & 
\ \mathcal{A}_2({\bf m},{\bf n};\{i,\alpha\},\{j,\beta\}) 
\end{align}
The matrix elements of quartic  operators of the 
form $a^{\dagger}_{\alpha}({\bf  p}_i)a^{\dagger}_{\beta}({\bf  p}_j)a_{\epsilon}({\bf  p}_k)a_{\zeta}({\bf  p}_l)$ are
\begin{align}\label{eq:quar_me}
\frac{1}{V^2} & \ \langle {\bf m } | a^{\dagger}_{\alpha}({\bf {p}}_i)~a^{\dagger}_{\beta}({\bf {p}}_j)~a_{\epsilon}({\bf {p}}_k)~a_{\zeta}({\bf {p}}_l)~| {\bf n} \rangle \nonumber\\
    = & \ 
    \Bigl(f_{{\bf m},i,\alpha} ~\delta_{m_{i,\alpha},1}\Bigr) 
    \Bigl( f_{{\bf m}^{[i \alpha]},j,\beta} ~\delta_{({\bf m}^{[i \alpha]})_{j,\beta},1}\Bigr)\nonumber\\
    \times& \Bigl(f_{{\bf n}^{[l \zeta]},k,\epsilon} ~\delta_{ ({\bf n}^{[l \zeta]})_{k,\epsilon},1} \Bigr)
    \Bigl( f_{{\bf n},l,\zeta} ~\delta_{n_{l,\zeta},1}\Bigr) \langle {\bf m}^{[i \alpha][j \beta]} | {\bf n}^{[l \zeta] [k \epsilon]} \rangle 
\nonumber \\
\equiv & 
\ \mathcal{A}_4({\bf m},{\bf n};\{i,\alpha\},\{j,\beta\},\{k,\epsilon\},\{l,\zeta\}). 
\end{align}
where 
\begin{equation}
{\bf n}^{[i\alpha][j \beta]} = {\bf n}^{[i \alpha]} ~~\mathrm{with}~~ n_{j,\beta}\rightarrow 0~. 
\end{equation} 

With Eq.~(\ref{eq:quad_me},\ref{eq:quar_me}), the Hamiltonian's matrix elements can be written in the following way. 
First, the kinetic energy, including the vacuum mixing terms, reads
\begin{widetext}
\begin{align}
    \langle {\bf m} | H_{\mathrm{Kin}} | {\bf n}\rangle
    &=  \sum_{i}  \bigg[  \left( |{\bf p}_i| +  \frac{m_1^2 + m_2^2 - \cos 2 \theta \, \delta m^2}{4 |{\bf p}_i|}  \right)  \mathcal{A}_2({\bf m},{\bf n};\{i,e\},\{i,e\}) 
\nonumber \\
&+ 
 \left( |{\bf p}_i| +  \frac{m_1^2 + m_2^2 + \cos 2 \theta \, \delta m^2}{4 |{\bf p}_i|}  \right)  
 \mathcal{A}_2({\bf m},{\bf n};\{i,\mu\},\{i,\mu\})  
\nonumber\\
&+ \frac{\sin 2 \theta \ \delta m^2}{4 |{\bf p}_i|} 
\Big(  \mathcal{A}_2({\bf m},{\bf n};\{i,e\},\{i,\mu\})  +  \mathcal{A}_2({\bf m},{\bf n};\{i,\mu\},\{i,e\})   \Big)
\bigg]~\text.
\end{align}

The matrix elements of the  normal-ordered interaction Hamiltonian can be written using Eq.~(\ref{eq:quar_me}).
For the full Hamiltonian including non-forward terms,  one finds 
\begin{align}
\langle {\bf m} | H_{\nu\nu}^{(F)} | {\bf n}\rangle 
&= - \frac{1}{V}
\frac{G_F}{2\sqrt{2}} \sum_{\alpha,\beta=e,\mu}~\sum_{i,j,k,l} 
\delta_{{\bf z}_{p_i} + {\bf  z}_{p_j} -{\bf  z}_{p_k} -{\bf z}_{p_l},{\bf 0}} \ \ g({\bf p}_i,{\bf p}_k,{\bf p}_j,{\bf p}_l)\nonumber\\
&\;\;\;\;\; \times \Big[\mathcal{A}_4({\bf m},{\bf n};\{i,\alpha\},\{j,\beta\},\{k,\alpha\},\{l,\beta\})+\mathcal{A}_4({\bf m},{\bf n};\{i,\alpha\},\{j,\beta\},\{k,\beta\},\{l,\alpha\}) \Big]\;\text.
\label{eq:Hfull1}
\end{align}
Finally, restricting summation over $k,l$ in Eq.~(\ref{eq:Hfull1}) to the  forward limit ($k=i$ or $l=i$), we obtain 
the usual truncated (`forward scattering') Hamiltonian, 
\begin{align}\label{eq:Hfwd1}
&\langle {\bf m} | H_{\nu\nu}^{(T)} | {\bf n}\rangle \nonumber\\
=& -  \frac{1}{V}\frac{G_F}{\sqrt{2}} \sum_{\alpha,\beta=e,\mu}\sum_{i,j} \left( 1-{\bf \hat{p}}_i\cdot {\bf \hat{p}}_j \right)\Big[\mathcal{A}_4({\bf m},{\bf n};\{i,\alpha\},\{j,\beta\},\{i,\alpha\},\{j,\beta\})+\mathcal{A}_4({\bf m},{\bf n};\{i,\alpha\},\{j,\beta\},\{i,\beta\},\{j,\alpha\}) \Big]
\end{align}
\end{widetext}
where the summations of $i,j$ run over all momentum modes.


\section{Reduced density matrices and  entanglement entropy}\label{sec:rhoands}

The full many-body system described by the state  $|\Psi  \rangle$ in Eq.~(\ref{eq:MBstate}) 
can be partitioned into two  subsystems in various ways, and 
the corresponding entanglement can be studied. 
In this context, the key object is the reduced density matrix, obtained by tracing over one of the 
two subsystems, starting from the full description of the state given by the density operator 
\begin{equation}
    \rho  = |\Psi  \rangle \langle \Psi |  = 
    \sum_{{\bf n},{\bf m}}
    \ c_{{\bf n}}  c_{{\bf m}}^*  \ |  {\bf n} \rangle \langle {\bf m} | ~.
\end{equation}

A central object in our study is the one-body reduced density matrix, 
that corresponds to partitioning the system into one particle versus $N-1$ particles 
and tracing  over $N-1$ particle states.  
This object is of great interest because 
(i) Neutrino measurements involve interactions of single neutrinos, hence 
knowledge of the one-body density matrix allows one to predict all observables of interest; 
 (ii) The von Neumann entropy computed in terms of the one-body 
 density matrix  quantifies the degree of entanglement 
of a single neutrino with the other $N-1$; (iii) the quantum kinetic equations (QKEs) 
are evolution equations for the one-body reduced density matrix. Therefore, 
 the one-body reduced density matrix provides the common ground on which 
 one can study and compare QKEs and many-body approaches.

Denoting the single-particle states  by $|\psi_i \rangle = | {\bf p}_{k_i}, \alpha_{f_i} \rangle$ (with 
$\alpha_{f_i} \in \{e, \mu \}$),  using the notation introduced in 
Section~\ref{sect:Fock} the one-body reduced density matrix takes the form
\begin{eqnarray}
    \rho^{(1)} &=& \sum_{i,j}   \, | \psi_i \rangle \langle \psi_j | \ \rho^{(1)}_{ij}~, 
    \\
    \rho^{(1)}_{ij} &=& \frac{1}{N} \sum_{{\bf n},{\bf m}} \ c_{{\bf n}}  c_{{\bf m}}^* \, \delta_{n_i 1} \delta_{m_j 1} 
    \, f_{{\bf n},i} f_{{\bf m},j} \, \delta_{{\bf  n}^{[i]}, {\bf m}^{[j]}} ~.
\end{eqnarray}
An alternative, very useful expression for the elements of the one-body reduced density matrix in terms expectation values 
of creation and annihilation operators of single-particle states is given by
\begin{equation}
      \rho^{(1)}_{ij} = \frac{1}{N}  \ \langle \Psi | \frac{a_j^\dagger}{\sqrt{V}} \frac{a_i}{\sqrt{V}} | \Psi \rangle~. 
\end{equation}
The  von Neumann entropy computed with $\rho^{(1)}$, 
\begin{equation}
    S (\rho^{(1)})= - {\rm Tr} \left( \rho^{(1)} \ \log \rho^{(1)} \right)~, 
\end{equation}
provides a measure of the entanglement between a single particle and the rest of the system~\cite{Nielsen:2012yss}.

Invariance under translations and orthogonality of single-particle momentum eigenstates implies that 
$\rho^{(1)}$ has a block structure on orthogonal single-particle subspaces labeled by three-momentum:
\begin{eqnarray}
    \rho^{(1)} &=&  \sum_{i=1}^{k} \ \sum_{\alpha,\beta  \in \{e, \mu\}}  
    \, | {\bf p}_i, \alpha \rangle \langle {\bf p}_i, \beta | \ \, \rho^{(1)}_{\alpha \beta} ({\bf p}_i)~, 
    \\
    \rho^{(1)}_{\alpha \beta} ({\bf p}) &=& \frac{1}{N}  \ \langle \Psi | \frac{a_\beta^\dagger ({\bf p})}{\sqrt{V}} \frac{a_\alpha ({\bf p})}{\sqrt{V}} | \Psi \rangle~. 
\label{eq:1bdm1}
\end{eqnarray}
Up to a normalization,  $\rho^{(1)}_{\alpha \beta} ({\bf p})$ is the dynamical quantity appearing in the QKEs. 
The diagonal entries are positive definite and represent the occupation numbers of 
electron and muon neutrinos in momentum ${\bf p}$, normalized to $N$. 
For each momentum ${\bf p}_i$, it is convenient to define 
\begin{equation}
    N^\pm_i  = \rho^{(1)}_{ee} ({\bf p}_i) \pm \rho^{(1)}_{\mu \mu} ({\bf p}_i) ~. 
    \label{eq:Nipm}
\end{equation}
$N^+_i$ is the total occupation number in momentum ${\bf p}_i$ (normalized to $N$), and it 
is very useful to study the kinetic properties of the state. 
On the other hand, $N^-_i$ characterizes the flavor content of the state in momentum ${\bf p}_i$.
In what follows, among other things we will use the time dependence of $N_i^\pm$ to identify the 
time scales of flavor and kinetic evolution. 

$\rho^{(1)}_{\alpha \beta} ({\bf p}_i)$ is not a density matrix:  its trace over flavor indices 
is $N_i^+/N$, where $N_i^+$  
is the total occupation number of momentum bin ${\bf p}_i$, irrespective of flavor (see Eq.~(\ref{eq:Nipm})). 
When $N_i^+ \neq 0$, rescaling by $N_i^+/N$, we can define  density matrices 
in flavor space for each momentum bin, 
\begin{eqnarray}
    \bar \rho^{(i)} &=& \sum_{\alpha,\beta  \in \{e, \mu\}}  
    \, | {\bf p}_i, \alpha \rangle \langle {\bf p}_i, \beta | \ \, 
    \bar \rho^{(i)}_{\alpha \beta} ({\bf p}_i)~, 
    \\
      \bar \rho^{(i)}_{\alpha \beta} ({\bf p}_i) &=&
\frac{N}{N_i^+} \ \rho^{(1)}_{\alpha \beta} ({\bf p})~, 
\end{eqnarray}
in terms of which the full one-body density matrix reads~\footnote{Note that 
when $N_i^+ \to 0$, all entries of $\rho^{(i)}_{\alpha \beta} ({\bf p}_i)$ vanish, 
so that there is no singularity in the definition of $\bar \rho^{(i)}$. 
Moreover,  in the limit $N_i^+ \to 0$ one also has
$N_i^+ \bar \rho^{(i)} \to 0$, hence no contribution to $\rho^{(1)}$ from 
momentum bin ${\bf p}_i$.}:
\begin{equation}
    \rho^{(1)} = \sum_{i=1}^{k}  \ \frac{N_i^+}{N}   \ \bar \rho^{(i)} ~.
\end{equation}

Finally, the von Neumann entropy of $\rho^{(1)}$ can be written as 
\begin{equation}
    S (  \rho^{(1)} ) = - \sum_{i} \frac{N_i^+}{N} \log \frac{N_i^+}{N}    + \sum_i \frac{N_i^+}{N}  \ S (\bar \rho^{(i)})~. 
    \label{eq:ee1}
\end{equation}
This expression shows that the entanglement entropy of a single neutrino with the rest of the system 
can arise from entanglement in both momentum (first term above) and flavor (second term above).

To make the above statements more precise, 
we can further trace the one-body density matrix over flavor or momentum degrees of freedom. 
Tracing over flavor, we get
\begin{eqnarray}
\rho^{(1),K} &=& \sum_{i=1}^{k} \ \frac{N_i^+}{N}  \ | {\bf p}_i \rangle \langle {\bf p}_i|~, 
\\
S (\rho^{(1),K}) &=& - \sum_{i} \frac{N_i^+}{N} \log \frac{N_i^+}{N} ~. \label{eq:ek1}
\end{eqnarray}
Similarly, tracing over momenta gives 
\begin{eqnarray}
\rho^{(1),F}_{\alpha \beta} &=& \sum_{i=1}^{k} \ \rho^{(1)}_{\alpha \beta} ({\bf p}_i)~, 
\end{eqnarray}
with $\rho^{(1)}_{\alpha \beta} ({\bf p}_i)$ defined in Eq.~(\ref{eq:1bdm1}).
For an initial state containing  $N_e$ electron neutrinos and  $N_\mu$  muon neutrinos across all momentum modes, 
one finds $\rho^{(1),F} = diag (N_e/N, N_\mu/N)$.

The time over which  $S (\rho^{(1)})$,  $S (\rho^{(1),K})$,  $S (\rho^{(1),F})$, 
 $S (\bar \rho^{(i)})$ reach the first maximum are proxy timescales for global, kinetic, and flavor equilibration. 
 We will illustrate these points in Section~\ref{sect:grid}.


\section{Setup for hot and dense media}
\label{sect:setup}

In this section, we discuss the energy scales 
characterizing the dynamics of neutrinos 
in situations of astrophysical interest  
and  define a rescaled, dimensionless Hamiltonian suitable for computational implementation. 
In situations of astrophysical interest, such as just below the decoupling region in a supernova, the initial state of the neutrinos is not too far from equilibrium. 
Therefore,  it makes sense to introduce a notion of near-equilibrium distribution and 
temperature, which characterizes the typical scale of the neutrino's momenta. 
In equilibrium, $N$, $T$, and $V$ are related by 
$N/V =  (3\zeta(3) T^3)/(4\pi^2)$. 
In our near-equilibrium situation, we assume that the above relation 
is approximately  valid and assume the scaling $1/V \sim T^3/N$ to estimate the relative 
size of the various contributions to the Hamiltonian~\footnote{For a temperature of $1$ MeV, we obtain 
$V \sim 10^8 N \,\,\text{fm}^3$.} 

When the temperature of the system is order MeV, 
widely separated  energy scales enter the Hamiltonian, that is, 
\be \label{eq:hierarchy}
| {\bf p}| \sim T  \gg   
G_F T^3
\gg  \delta m^2/T~.
\ee
The scales $T$ and 
 $G_F T^3$ differ by about ten orders of magnitude. 
As discussed below, this wide separation effectively 
removes the effect of self-interactions 
connecting neutrino pairs with different kinetic energy, i.e. with 
$|{\bf p}_i| + |{\bf p}_j| \neq |{\bf p}_i^\prime| + |{\bf p}_j^\prime|$. 
On the other hand, $G_F T^3$  and $\delta m^2/T$ differ by two-to-three orders of magnitude 
depending on the magnitude of the mass splitting used. These two scales together control the flavor evolution of the system. 

To make the interplay between the three parameters more explicit, we rescale the Hamiltonian for the rest of the discussion. 
We take as the unit of energy (and inverse time) the quantity 
${\cal E} \equiv G_F/(\sqrt{2}V)$, and introduce the dimensionless parameters 
\begin{eqnarray}
    \bar T &=&
     \frac{T}{\cal E} \sim 10^{10}
    \\
    \bar \omega &=& 
    \frac{\delta m^2}{4T {\cal E}}  \sim 10^{-3} - 10^{-2}
    \\
    |\tilde {\bf p} |&=& \frac{|{\bf p}|}{T} \sim  O(1).
\end{eqnarray}
The Hamiltonian of the neutrinos can now be written 
in terms of dimensionless operators in the following way: 
\begin{align}
    H^{(F/T)} &= {\cal E} \ \left( \bar H_{\rm Kin} + \bar H^{(F/T)}_{\nu\nu} \right) \label{eq:Htoy}\\
    \bar H_{\rm Kin} &= \frac{1}{V}\sum_{i=1}^k \left( \bar T |\tilde{\bf p}_i| - \frac{\bar \omega \cos 2\theta}{|\tilde{\bf p}_i|} \right) a^{\dagger}_e({\bf p}_i)a_e({\bf p}_i) \nonumber\\
    &~~~~~~~~~~~+ \left( \bar T |\tilde{\bf p}_i| + \frac{\bar \omega \cos 2\theta}{|\tilde{\bf p}_i|} \right)a^{\dagger}_{\mu}({\bf p}_i)a_{\mu}({\bf p}_i) \nonumber\\
     &~~~~~~~~~~~+\frac{\bar \omega \sin 2\theta}{|\tilde{\bf p}_i|} \Big[ a^{\dagger}_e({\bf p}_i)a_{\mu}({\bf p}_i) + a^{\dagger}_{\mu}({\bf p}_i)a_{e}({\bf p}_i) \Big] \nonumber\\
    \bar  H_{\nu\nu}^{(F/T)} &= - \frac{1}{V^2}\sum_{(i,j,k,l)} g({\bf p}_i,{\bf p}_k,{\bf p}_j,{\bf p}_l) \nonumber\\
    &~~~~~~~~~~~~~~ \times a^{\dagger}_{\alpha}({\bf p}_i)a^{\dagger}_{\beta}({\bf p}_j)a_{\alpha}({\bf p}_k)a_{\beta}({\bf p}_l)~. \nonumber
\end{align}
Note that we have dropped the terms proportional to $(m_1^2 + m_2^2)/|{\bf p}_i|$ in the diagonal part of the vacuum term $\bar H_{\rm Kin}$, as they are always subleading compared to the term proportional to $|{\bf p}_i| \to \bar T |\tilde{\bf p}_i|$. The summation $(i,j,k,l)$ in $\bar H_{\nu\nu}$ denotes all pairs that conserve 3-momentum for the full Hamiltonian $H^{(F)}$. In the truncated case $\bar H_{\nu\nu}^{(T)}$, we additionally impose that ${\bf p}_i={\bf p}_k$ and ${\bf p}_j={\bf p}_l$, or ${\bf p}_i={\bf p}_l$ and ${\bf p}_j={\bf p}_k$.


\section{A toy problem with $N=2$}
\label{sect:toy}

In this section, we explore the impact of the non-forward interaction 
on the evolution of flavor and momentum degrees of freedom   
in a dense neutrino system by inspecting a toy model. In the model, there are 4 momentum modes ($k=4$) for neutrinos to fill, and we consider states with two neutrinos ($N=2$). Therefore the dimension of the relevant Hilbert space is $d_{2,4}=28$. The momentum modes $\tilde {\bf p} = {\bf p}/T$ in this toy system are chosen as 
follows:
\begin{align}
    \tilde{\bf p}_1 &=(\sin\phi,~\cos\phi,~0)\nonumber\\
    \tilde{\bf p}_2 &=(\sqrt{r^2-\cos^2\phi},~-\cos\phi,~0)\nonumber\\
    \tilde{\bf p}_3 &=(\sin\phi,~-\cos\phi-\varepsilon,~0)\nonumber\\
    \tilde{\bf p}_4 &= (\sqrt{r^2-\cos^2\phi},~\cos\phi+\varepsilon,~0)~, 
    \label{p}
\end{align}
with $\phi\in\left[0,\pi/2\right]$, $r>1$ and $\varepsilon\in\mathbb{R}$. In this system, ${\bf p}_1+{\bf p}_2={\bf p}_3+{\bf p}_4$ for any $r, \phi$ and $\varepsilon$. Therefore the quartic terms in $\bar H_{\nu\nu}$ involving creation of ${\bf p}_1, {\bf p}_2$ (${\bf p}_3, {\bf p}_4$) and annihilation of ${\bf p}_3, {\bf p}_4$ (${\bf p}_1, {\bf p}_2$) will have non-zero matrix elements. Note also that $|{\bf p}_1|+|{\bf p}_2| = |{\bf p}_3|+|{\bf p}_4|$ only when $\varepsilon=0$. One important finding in this section is that the non-forward interaction  significantly affects the time evolution only when $\varepsilon  \lesssim 1/\bar T$, i.e. when the difference in kinetic energy of the two pairs is on the order of or smaller than 
potential energy due to the self-interactions. 

The Hamiltonian is block diagonal in the basis of Eq.~(\ref{eq:basis}) since it commutes 
with the total momentum operator and hence connects 
states with the same total momentum. Among the 28 basis states, 4 states have two neutrinos with the same momentum and different flavor. These 4 states have total momentum of $2{\bf p}_i$, and they each form a $1\times 1$ block. Second, for the blocks with a total momentum of ${\bf p}_i+{\bf p}_j $ $(i\neq j)$, there are 4 such states in each block due to the choice of flavor for each neutrino. Therefore there are 
six 
$4\times4$ blocks with those total momenta. Finally, since ${\bf p}_1+{\bf p}_2={\bf p}_3+{\bf p}_4$, blocks with total momenta ${\bf p}_1+{\bf p}_2$ and ${\bf p}_3+{\bf p}_4$ are connected via $\bar H_{\nu\nu}$ for the full Hamiltonian. In summary, for the only-forward Hamiltonian, there are four $1\times1$ blocks and six $4\times4$ blocks, while for the full Hamiltonian, there are four $1\times1$ blocks, four  $4\times4$ blocks, and one $8\times8$ block connecting states with ${\bf p}_1+{\bf p}_2$ and ${\bf p}_3+{\bf p}_4$. 

Let us take a closer look at the $8\times8$ block. We order the eight basis vectors $|v_1 \rangle ... |v_8 \rangle$  as follows:
\begin{eqnarray}
 V  |v_1\rangle &=
   a^{\dagger}_e({\bf p}_1)a^{\dagger}_e({\bf p}_2) |0\rangle\;,~~~~ V  |v_5\rangle &=  a^{\dagger}_e({\bf p}_3)a^{\dagger}_e({\bf p}_4) |0\rangle\; \nonumber\\
    V |v_2\rangle &= 
    a^{\dagger}_e({\bf p}_1)a^{\dagger}_\mu ({\bf p}_2) |0\rangle\;,~~~~ V|v_6\rangle &= a^{\dagger}_e({\bf p}_3)a^{\dagger}_\mu ({\bf p}_4) |0\rangle\; \nonumber\\
   V |v_3\rangle &= a^{\dagger}_{\mu}({\bf p}_1)a^{\dagger}_e({\bf p}_2) |0\rangle\;,~~~~ V |v_7\rangle &= a^{\dagger}_{\mu}({\bf p}_3)a^{\dagger}_e({\bf p}_4) |0\rangle\; \nonumber\\
   V |v_4\rangle &= a^{\dagger}_{\mu}({\bf p}_1)a^{\dagger}_{\mu}({\bf p}_2) |0\rangle\;,~~~~  V |v_8\rangle &= a^{\dagger}_{\mu}({\bf p}_3)a^{\dagger}_{\mu}({\bf p}_4) |0\rangle\;~. \nonumber
\end{eqnarray}
To write down the matrix elements of the Hamiltonian in these basis states, we first introduce shorthand notations:
\begin{align}
    D_{ij}(\pm_i,\pm_j) &= \bar T (|\tilde {\bf p}_i|+|\tilde {\bf p}_j|) \pm_i \frac{\bar \omega \cos 2\theta}{|\tilde {\bf p}_i|} \pm_j \frac{\bar \omega \cos 2\theta}{|\tilde {\bf p}_j|}\\
    \bar \omega_i &= \frac{\bar \omega \sin 2\theta}{| \tilde{\bf p}_i|}
\end{align}
for the kinetic parts and
\begin{align}
    f_{ij} &=  f({\bf p}_{i},{\bf p}_{j})  \label{eq:fij}\\
    M & = \begin{pmatrix}
    4 & 0 & 0 & 0\\
    0 & 2 & 2 & 0 \\
    0 & 2 & 2 & 0 \\
    0 & 0 & 0 & 4
    \end{pmatrix}
\end{align}
for the interactions, with the function $f({\bf p}_{i},{\bf p}_{i})$ in Eq.~(\ref{eq:fij}) defined in Eq.~({\ref{eq:f}}). With these notations, 
the kinetic part of the Hamiltonian,  
$\bar H_{\rm Kin}$, including vacuum mixing, takes the form  
\begin{widetext}
  \begin{equation}
   \bar H_{\rm Kin} = \begin{pmatrix}
D_{12}(-,-) & \bar \omega_2 & \bar \omega_1 & 0 & 0 & 0 & 0 & 0\\
\bar \omega_2 & D_{12}(-,+) & 0 & \bar \omega_1 & 0 & 0 & 0 & 0\\
\bar \omega_1 & 0 &D_{12}(+,-)& \bar \omega_2 & 0 & 0 & 0 & 0\\
0 & \bar \omega_1 & \bar \omega_2 & D_{12}(+,+)& 0 & 0 & 0 & 0\\
0 & 0 & 0 & 0 & D_{34}(-,-) & \bar \omega_4 & \bar \omega_3 & 0\\
0 & 0 & 0 & 0 & \bar \omega_4 & D_{34}(-,+) & 0 & \bar \omega_3\\
0 & 0 & 0 & 0 & \bar \omega_3 & 0 & D_{34}(+,-) & \bar \omega_4\\
0 & 0 & 0 & 0 & 0 & \bar \omega_3 & \bar \omega_4 & D_{34}(+,+)
\end{pmatrix} ~.
\end{equation}  
\end{widetext}
The interaction terms read
\begin{equation}
   \bar H^{(T)}_{\nu\nu} = \begin{pmatrix}
f^{\dagger}_{12}f_{12}  M &0 \\
0 & f^{\dagger}_{34}f_{34} M
\end{pmatrix} 
\end{equation}
and 
\begin{equation}
   \bar H^{(F)}_{\nu\nu} = \begin{pmatrix}
f^{\dagger}_{12}f_{12} M & f^{\dagger}_{12}f_{34}  M \\
f^{\dagger}_{34}f_{12}  M & f^{\dagger}_{34}f_{34}  M
\end{pmatrix}.
\end{equation}

\begin{figure*}
    \centering
    \includegraphics[width=0.47\textwidth]{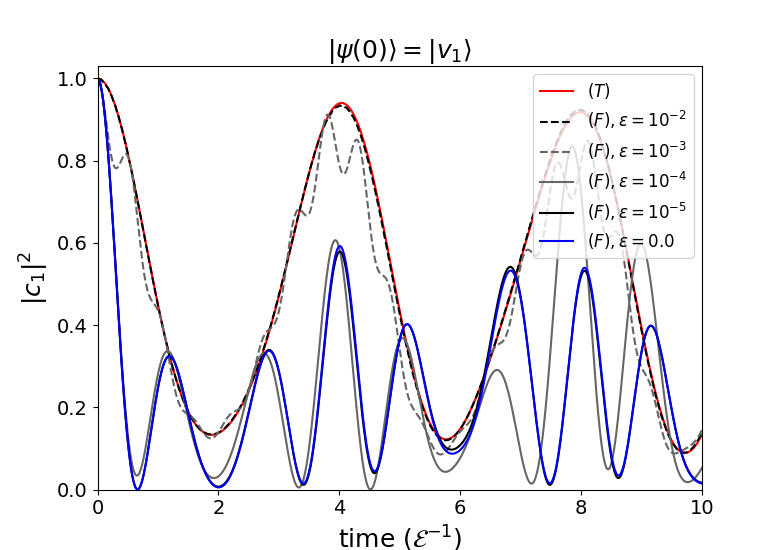} 
    \hspace{.5cm}
    \includegraphics[width=0.47\textwidth]{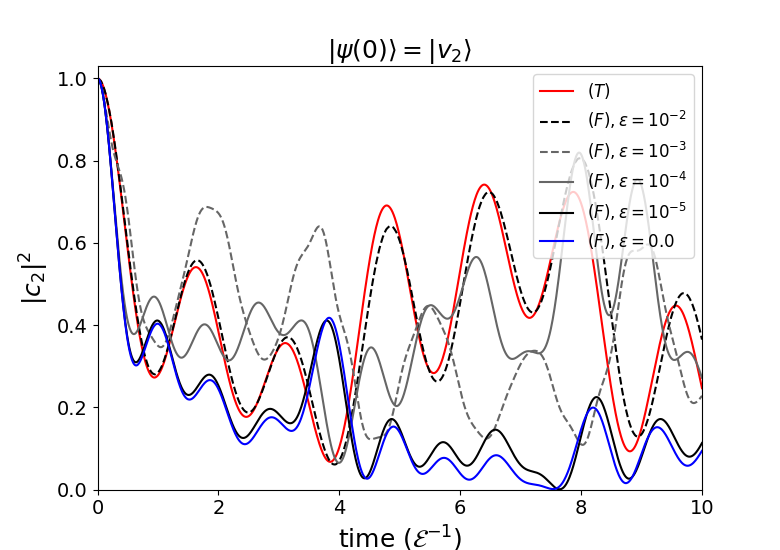} 
    \caption{ \label{fig:toy_kec}
    Loschmidt echo $|\langle \psi(0)| \psi (t) \rangle|^2$ in the two-neutrino toy model, 
    for  $|\psi (0) \rangle = | v_1 \rangle$   (left panel) and  $|\psi (0) \rangle = | v_2 \rangle$   (right panel). 
    In both cases the Hamiltonian parameters are $\bar T=10^4$,  $\bar\omega=1$, and $\sin 2\theta=0.8$. The momentum parameters are $r=2.0$, $\phi=\pi/4$ and $\varepsilon=0, 10^{-2} - 10^{-5}$.}
\end{figure*}

One thing to note about the eigenvalue spectrum of $\bar H_{\nu\nu}$ is that 5  of them are zero and three take the same value.
(When considering the total Hamiltonian, the vacuum mixing terms in $\bar H_{\rm Kin}$ lift this degeneracy.)
This is due to the fact that  
each block is given by the $4\times4$ matrix $M$ multiplied by the products of two $f_{ab}$ factors,
corresponding to the incoming and outgoing momentum pairs. 
As a result, the first 4 columns of $\bar H_{\nu\nu}$ and the last 4 are linearly dependent. 
Together with the structure of  $M$, this implies that 
the rank of the matrix $\bar H_{\nu\nu}$ is 3. This is a special feature seen only in 2-neutrino systems.

In contrast to the above point, there is an important  general 
feature of the problem that emerges from the analysis of this toy model. 
Given that in situations of (astro)physical interest $\bar T/\bar \omega > 10^{10}$, 
 when  $|\tilde {\bf p}_1|+|\tilde {\bf p}_2|$ and $|\tilde {\bf p}_3|+|\tilde {\bf p}_4|$ differ by $O(1)$, transitions between the blocks with the total momentum of ${\bf p}_1+{\bf p}_2$ and ${\bf p}_3+{\bf p}_4$ caused by $H^{(F)}$ become negligible. The two blocks are dynamically decoupled, similarly to what happens in any two-level quantum system 
when the off-diagonal mixing term is much smaller than the  unperturbed level splitting.  
In this regime,   the evolution is effectively controlled by the truncated Hamiltonian $\bar H^{(T)}$.   
Only when the difference between $|\tilde {\bf p}_1|+|\tilde {\bf p}_2|$ and $|\tilde {\bf p}_3|+|\tilde {\bf p}_4|$ is of similar magnitude as  $|f_{12}^{\dagger}f_{34}|/\bar T$, the neutrino-neutrino non-forward interactions contribute to the dynamics. To test this observation, we analyse  the time evolution of the toy model with $\bar T=10^4$,  $\bar\omega=1$, and $\sin 2\theta=0.8$ as we vary the 
parameter $\varepsilon$ that controls the ``kinetic energy conservation" condition 
through $ |\tilde {\bf p}_3|+|\tilde {\bf p}_4| =  |\tilde {\bf p}_1|+|\tilde {\bf p}_2|+ O(\varepsilon)$.
 We fix the momentum parameters to be $r=2.0$ and $\phi=\pi/4$ 
and study 
\begin{equation}
|\psi (t) \rangle = \sum_{i=1}^8  \ c_i (t)  \ |v_i \rangle ~.
\end{equation}
For the purpose of illustration, we show results for initial 
states $|\psi (0) \rangle = | v_1 \rangle$ (two electron neutrinos with momentum ${\bf p}_1$ and  ${\bf p}_2$) and 
$|\psi (0) \rangle = | v_2 \rangle$
(electron neutrino with momentum ${\bf p}_1$ and  muon neutrino with momentum ${\bf p}_2$). 
In Fig.~\ref{fig:toy_kec} we plot the Loschmidt echo $|\langle \psi(0)| \psi (t) \rangle|^2$
as a function of time   
and compare it to the time evolution obtained the truncated (``forward") Hamiltonian.
As expected,  $|\langle \psi(0)| \psi (t) \rangle|^2$
starts to significantly deviate from the truncated behavior around $\varepsilon \sim 1/\bar T = 10^{-4}$ and the effect of non-forward scattering increases as $\varepsilon$ decreases.   For $\varepsilon \sim 10^{-5} \sim 0.1/\bar T$ or smaller,  the evolution 
becomes essentially indistinguishable from the $\varepsilon=0$ case.

This dynamical pairwise kinetic energy conservation limits     
the number of relevant  terms in the full Hamiltonian $H^{(F)}$.  
This is still much less restrictive than the `forward' kinematics enforced 
in the truncated Hamiltonian $H^{(T)}$. 
In fact,   for each pair of momenta ${\bf p}$ and ${\bf q}$,    
three-momentum and kinetic energy conservation open up 
an infinite set of momentum pairs ${\bf p}^\prime$ and ${\bf q}^\prime$  
(parameterized by two angles as shown in Appendix~\ref{sect:appendix2}) that contribute to the evolution, 
to be contrasted to just one option in `forward' kinematics. 
The pairwise kinetic energy conditions becomes very restrictive 
only if one considers a 1-dimensional set up. 
In this case, the discussion below  Eq.~(\ref{eq:planar}) implies that  
$|{\bf p}_z| + |{\bf q}_z| = |{\bf p^\prime}_z |+ |{\bf q^\prime}_z |$
only holds if  ${\bf p}^\prime_z = {\bf p}_z$ or ${\bf p}^\prime_z = {\bf q}_z$. 
In other words, pairwise kinetic energy condition enforces forward or exchange kinematics 
and the evolution with the full and truncated Hamiltonian effectively coincide.

\begin{figure}
        \centering
    \includegraphics[width=0.5\textwidth]{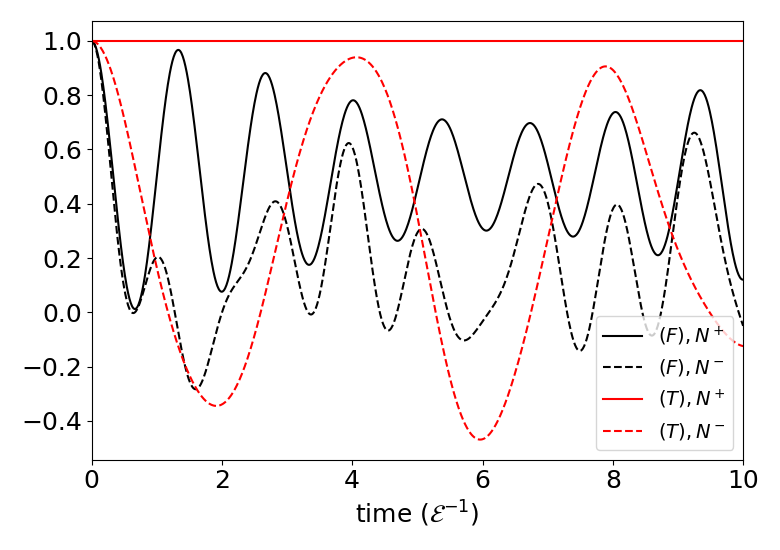}
    \caption{Expectation values of $N_1^{+}$ and $N_1^-$ for the mode ${\bf p}_1$ over time for the full and truncated Hamiltonian. The Hamiltonian parameters are chosen as $\bar T=10^4$,  $\bar\omega=1.0$, and $\sin 2\theta=0.8$. The momentum parameters are $r=2.0, \phi=\pi/4$ and $\varepsilon=0.0$. The initial state is $v_1$, i.e., neutrinos with flavor $e$ and momentum ${\bf p}_1, {\bf p}_2$. \label{fig:toy_bin1}} 
\end{figure}

When kinetic energy conservation holds approximately, i.e. at $\varepsilon\sim0$, all eight states with total momentum of ${\bf p}_1+{\bf p}_2$ (or equivalently ${\bf p}_3+{\bf p}_4$) get non-zero amplitudes over time without suppression via $\bar T$. This spread of amplitude is governed by the interplay between the neutrino's self-interaction and their vacuum oscillation. One then naturally expect to see the effect of non-forward scattering in 
flavor equilibration and 
randomization of momenta. 
 We can see a hint of this even in this two-neutrino toy model. 
To study these phenomena,  in Fig.~\ref{fig:toy_bin1}, we compare the $N_1^+$ and $N_1^-$ 
(see definition in Eq.~(\ref{eq:Nipm})).
for the state evolved via either $H^{(F)}$ (in black) or $H^{(T)}$ (in red). Obviously, kinetic evolution occurs only with the full Hamiltonian --- 
the occupation number of a given momentum mode cannot change via forward scattering or exchange processes. 
This is evident in the constant solid red line showing the total occupation number of the momentum mode. Regarding flavor conversion, which occurs for both full and truncated set-ups, the evolution speed at initialization appears to be faster with the full Hamiltonian. This is correlated with the fact that the total occupation of the mode decreases quickly due to the non-forward interaction. Therefore, at least in this toy model, 
the time scales for kinetic evolution (and ultimately  thermalization)   and flavor evolution are  correlated.
We will further investigate the relation between kinetic and flavor equilibration in the next section.

\begin{figure}
        \centering
    \includegraphics[width=0.5\textwidth]{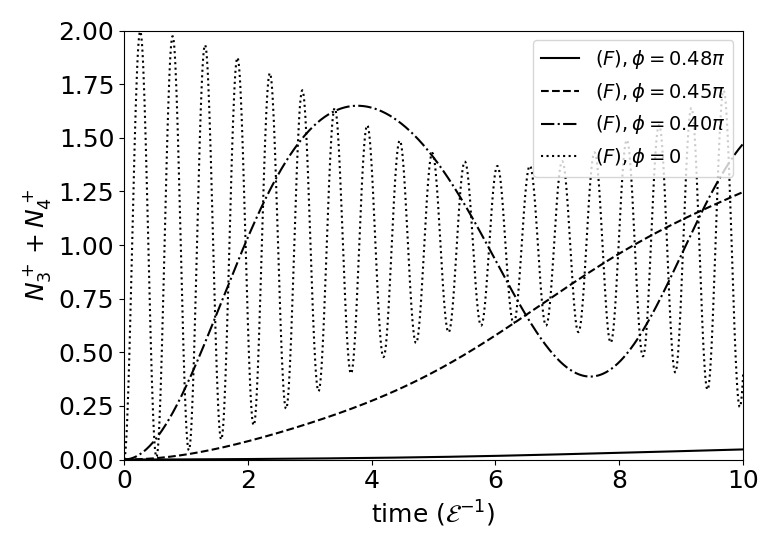}
    \caption{ Time evolution induced by the full Hamiltonian $H^{(F)}$ 
    for the sum of occupation numbers $N_3^+ + N_4^+$ of initially unoccupied momentum modes 
    ${\bf p}_3$ and ${\bf p}_4$  ($| \psi (0) \rangle =|v_1\rangle$). 
    The Hamiltonian parameters are $\bar T=10^4$,  $\bar\omega=1$, and $\sin 2\theta=0.8$. The momentum parameters are $r=2.0$ and $\varepsilon=0$ while we vary $\phi$. \label{fig:toy_g}} 
\end{figure}

We close this section by pointing out 
that the evolution time scales can be  affected by 
the angular factors 
$g ({\bf p^\prime}, {\bf p}, {\bf q^\prime}, {\bf q})$
in the Hamiltonian matrix elements, given in  Eq.~(\ref{eq:g}). 
From now onwards, we will refer to these as the ``$g$ factors". 
The magnitude of the $g$ factors  depends on the 
relative angles of the incoming and outgoing momentum modes, 
parameterized in this simple model by $\phi$ in Eq.~(\ref{p}).
In Fig.~\ref{fig:toy_g}, we plot the time dependence 
of the sum of occupation numbers 
$N_3^+ + N_4^+$ of initially unoccupied momentum modes  ($| \psi (0) \rangle =|v_1\rangle$). 
for different choices of the angle $\phi$ in Eq.~(\ref{p}). 
As $\phi$ approaches $\pi/2$, the angle between ${\bf p}_1$ and ${\bf p}_2$ decreases 
and the effect of non-forward scattering vanishes. 
This has physical implications  related to geometric effects, as 
neutrino crossing angles away from a source are geometrically suppressed. 
Besides this, even in absence of geometric suppression, 
this feature is expected  
slow-down the evolution  
time scales in simulations with small number of neutrinos 
and small number of available momentum modes. 
Some of these artifacts will appear in our discussion in the next Section.

In this section, we studied several key features of the full Hamiltonian in the two-neutrino toy model:
\begin{itemize}
\item Most notably, the hierarchy between the neutrino kinetic and potential energy ($\bar T \gg 1$) 
results in pair-wise kinetic energy conservation. 
Together with three-momentum conservation, for each pair of momenta ${\bf p}$ and ${\bf q}$ 
this still opens up an infinite set of momentum pairs ${\bf p}^\prime$ and ${\bf q}^\prime$   
that contribute to the evolution (see  Appendix~\ref{sect:appendix2}). 
\item 
Non-forward processes  induce kinetic (momentum) randomization  and  have the  potential  for accelerating  flavor evolution. 
\item  
The quantitative impact  of non-forward terms in the Hamiltonian depends on 
the magnitude of the $g$ factors, which decrease as the relative angle of two incoming momentum modes decreases.
\end{itemize}

 
\section{Two-dimensional models}
\label{sect:grid}

In this section, we study the time evolution of various neutrino systems under the full Hamiltonian $H^{(F)}$ and truncated Hamiltonian $H^{(T)}$ with momentum modes taken on a two-dimensional grid:
\begin{align}\label{eq:gridm}
 \tilde {\bf p} & \equiv {\bf p}/T  = \frac{2 \pi}{L T} \ {\bf z}  
 \nonumber \\
 {\bf z} & =  \{  (z_x,z_y); ~z_x, z_y \in \mathbb{Z}\} \nonumber\\
     &\mathrm{with} ~~~ 0 <| {\bf z} | \le z_{\mathrm{max}} ~~ \mathrm{and}~~ 0 < z_x \;\text.
\end{align}
We exclude the zero mode and introduce a ``UV'' cut-off $p_{\mathrm{max}}$ that sets the maximal magnitude on the momenta. 
Moreover, we only consider the modes with a positive $x$ component to mimic astrophysical situations 
in which there is a net flux of neutrinos.

With this setup, we first demonstrate that kinetic energy conserves pair-by-pair through the self-interactions and show that we can safely take this as an exact conservation law at typical temperature of  interest $T\gtrsim$ MeV. Given that, we impose the pair-wise kinetic energy conservation on both the Hilbert space and Hamiltonian, and explore the collective flavor evolution and kinetic (momentum) randomization in this system while varying the number of neutrinos $N=6, 8, 10$ and initial states. All codes developed for the numerical simulations performed in this section are available online\footnote{ \href{https://github.com/yukariyamauchi/neutrinos_beyond_fwd}{https://github.com/yukariyamauchi/neutrinos\_beyond\_fwd}}.


\subsection{Pair-wise kinetic energy conservation}\label{subsec:2dkec}
As discussed in Section~\ref{sect:toy}, 
when the typical magnitude of neutrino three-momenta 
(dictated by the temperature) is much larger than the neutrino self-interaction potential energy, 
out of all couplings contained in  $H_{\nu \nu}$, 
only the ones satisfying approximate pair-wise kinetic energy conservation 
are expected to affect the dynamics.   
Here we demonstrate this  within  the two-dimensional models specified by the grid of momentum modes in Eq.~(\ref{eq:gridm}). 
For this purpose, we simulate the time evolution numerically via exact diagonalization of the two Hamiltonians: the full Hamiltonian and the one with the kinetic energy conservation imposed in $H_{\nu\nu}$, which we denote as $H_{\nu \nu}^{(K)}$. Explicitly, $H_{\nu \nu}^{(K)}$ is 
\begin{align}
    \bar H^{(K)} &= \bar H_{\rm Kin} + \bar H^{(K)}_{\nu\nu}\nonumber\\
    \bar H^{(K)}_{\nu\nu} &= -\frac{1}{V^2}\sum_{{\bf p}_{i,j,k,l} \in \{{\bf p}\}} \Big[ a^{\dagger}_{\alpha}({\bf p}_i)a^{\dagger}_{\beta}({\bf p}_j)a_{\alpha}({\bf p}_k)a_{\beta}({\bf p}_l) \nonumber\\
    & \times \delta_{{\bf p}_i+{\bf p}_j,{\bf p}_k+{\bf p}_l} ~ \delta_{|{\bf p}_i|+|{\bf p}_j|,|{\bf p}_k|+|{\bf p}_l|} ~ g({\bf p}_i,{\bf p}_k,{\bf p}_j,{\bf p}_l)
\end{align}
Comparison with  Eq.~(\ref{eq:Htoy}) reveals that 
$\bar H^{(T)}_{\nu\nu} \subset \bar H^{(K)}_{\nu\nu} \subset \bar H_{\nu \nu}^{(F)}$. 

We set $z_{\mathrm{max}}=3$ for the grid of momenta, which gives 11 modes. The simulation is performed in the Hilbert space with the number of neutrinos $N=2$ and 4. Therefore the dimension of the Hilbert space is $d_{N,11}=231$ or 7315 for $N=2, 4$ respectively. We do not make a further truncation to the Hilbert space. In the following demonstrations, the Hamiltonian parameters are $\bar\omega=1.0$ and $\sin 2\theta=0.8$.
We take as  initial state at time $t=0$  a  superposition of all basis states with $N$ electron neutrinos,  
 assigning an equal amplitude to all such basis states: 
\begin{equation}
    |\psi(0)\rangle = \frac{1}{\sqrt{_{11}C_N}}
\sum_{{\bf n}}
    \delta_{N,\sum_{i=1}^{11} n_{i,e}} | {\bf n} \rangle \;\text,
\end{equation}
where $_nC_k$ denotes the binomial coefficient.

\begin{figure}
    \centering
        \includegraphics[width=0.5\textwidth]{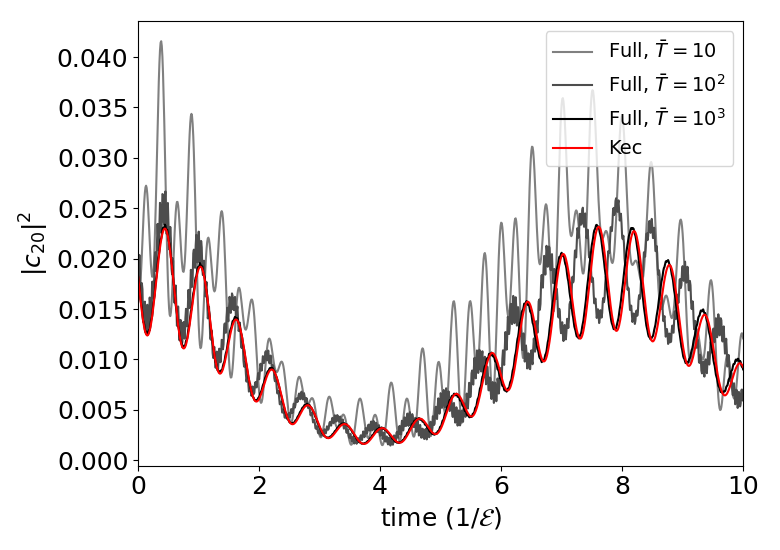}
    \caption{The squared modulus of the amplitude of the 20th basis state simulated with $H^{(F)}$ (denoted as ``Full'') and with $H^{(K)}$ (denoted as ``Kec''). The Hamiltonian parameters are $\bar\omega=1.0$ and $\sin 2\theta=0.8$. The model has $k=11$ momentum modes, and $N=2$ neutrinos.   Note that the black solid line is almost on top of the red solid line.\label{fig:tbar_Nn2_c14}}
\end{figure}

As an example, in the case of $N=2$ the squared modulus of the amplitude of the 20th basis state, which has electron neutrinos with momentum $(z_x, z_y)=(1,-2)$ and $(2,2)$ is shown in Fig.~\ref{fig:tbar_Nn2_c14}. When the temperature is low, the state can transition to other 2-neutrino states such as the one with momentum modes $(1,0)$ and $(2,0)$. However, as the temperature increases, such transitions are suppressed unless kinetic energy is conserved. In the infinite-temperature limit, the initial state can transition only to the states of neutrinos with momentum modes $(2,-2)$ and $(1,2)$ on the grid. Fig.~\ref{fig:tbar_Nn2_c14} shows that as the temperature $\bar T$ goes up, $|c_{20} (t)|^2$ converges to the time evolution with $H^{(K)}$. The difference between the amplitude from $H^{(F)}$ and $H^{(K)}$ is negligible at $\bar T=10^3$ (in black solid line).

\begin{figure}
    \centering
        \includegraphics[width=0.5\textwidth]{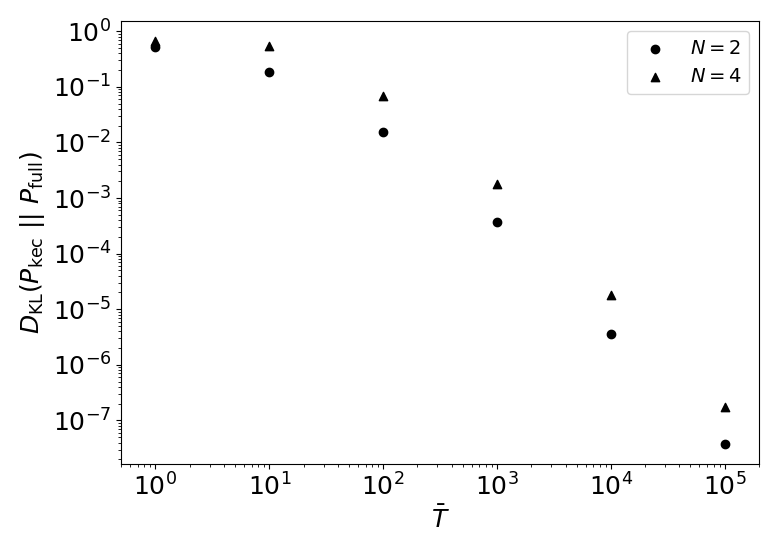}
    \caption{The Kullback–Leibler divergence of the two probability densities 
    $P_{\mathrm{kec}}(t = 10 / \mathcal{E})$ and $P_{\mathrm{full}}(t=10 /\mathcal{E})$  defined by the set of  $|c_n (t)|^2$ 
    evolved with $H^{(K)}_{\nu \nu}$ and $H^{(F)}_{\nu \nu}$, respectively. 
The Hamiltonian parameters are $\bar\omega=1.0$ and $\sin 2\theta=0.8$ while varying $\bar T$. The model has 11 momentum modes, and the number of neutrinos is 2 or 4.\label{fig:tbar}}
\end{figure}

To quantify more globally the difference between two simulations done with $H^{(F)}$ or $H^{(K)}$, we employ the Kullback–Leibler (KL) divergence~\cite{10.1214/aoms/1177729694} $D_{\mathrm{KL}}(P_{\mathrm{kec}}(t)~||~P_{\mathrm{full}}(t))$, 
where $P_{\mathrm{kec}}(t)$ and $P_{\mathrm{full}}(t)$ are probabilities defined by their corresponding $|c_n (t)|^2$.
In Fig.~\ref{fig:tbar} we show the KL divergence at time $t=10/\mathcal{E}$) for the $N=2,4$ systems while varying $\bar T$. The KL divergence decreases with temperature $\bar T$ according to a power law, thus offering a parametric evidence for the `dynamical' pair-wise 
kinetic energy conservation.

\subsection{Details of the simulated systems}

\begin{figure*}
    \centering
    \includegraphics[width=0.25\textwidth]{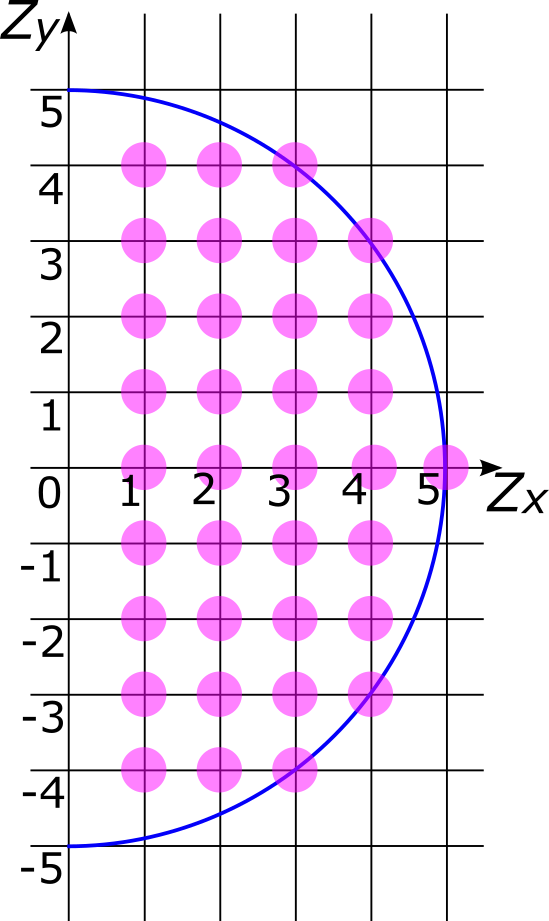}
    \hspace{0.05\textwidth}
    \includegraphics[width=0.5\textwidth]{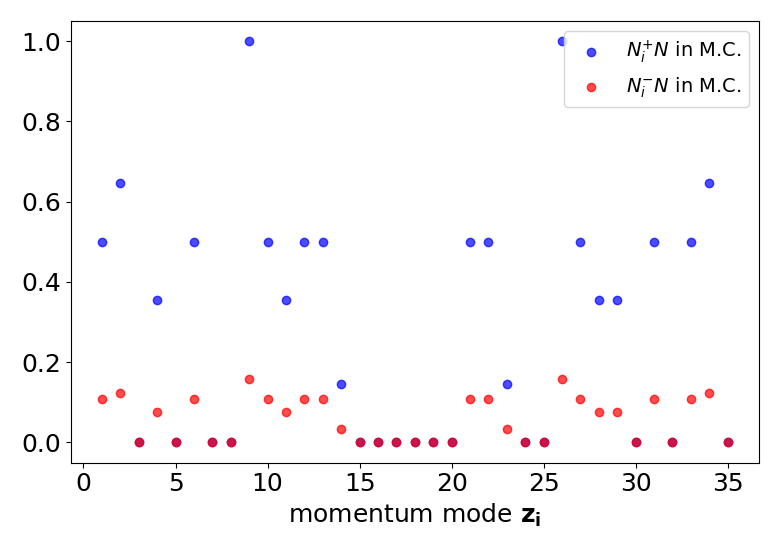}
    \caption{\label{fig:grid}  Left: 35 momentum modes are shown in pink circles on the grid with the maximal magnitude $z_{\mathrm{max}}=5$ (blue line). Right: The equilibrium expectation values of $N  N^{+/-}_i$ computed with Eq.~(\ref{eq:mc}) are shown in blue and red respectively for $\mathcal{H}_3$. }
\end{figure*}

In the rest of the section, we impose that pair-wise kinetic energy conservation holds exactly in the neutrino-neutrino self-interaction as in $H^{(K)}$. Additionally, we exclusively focus on the neutrino self-interaction and turn off the vacuum oscillations by setting $\bar\omega=0$. This condition allows us to decompose the entire Hilbert space into the subspaces with a fixed total momentum, kinetic energy, and particle numbers $N_e$ (electron neutrino) and $N_{\mu}$ (muon neutrino). Even within each subspace with these fixed quantities, there are multiple disconnected subspaces --- the matrix element of $H^{(F)}_{\nu\nu}$ is non-zero only when the two states have a pair of two momentum modes whose total momentum and kinetic energy are the same. On the momentum grid with $z_{max}=5$ with $k=35$ momentum modes as shown in the left panel of Fig.~{\ref{fig:grid}}, we focus on three such subspaces: $\mathcal{H}_1$ with  $N=6$, $\mathcal{H}_2$ with  $N=8$, and $\mathcal{H}_3$ with  $N=10$, where $N$ denotes the total number of neutrinos. We choose the number of electron neutrino to be $N_e = N/2+1$. Some defining quantities of the Hilbert spaces are summarised in Table.~\ref{table:h}. 

The basis states in these Hilbert spaces are enumerated by picking a ``reference basis state'' and listing all other basis states that can be reached from the reference state via repeatedly applying the self-interaction $\bar H^{(K)}_{\nu\nu}$. 
The reference state we used  for $\mathcal{H}_1$ is $|{\bf n}\rangle$ with
\begin{equation}
\begin{aligned}
    n_{ie} &= \begin{cases} 
                1 & i = 1,6,11,26\\
                0 & \mathrm{otherwise}
                \end{cases} \\
    n_{i\mu} &=\begin{cases} 
                1 & i = 9,34\\
                0 & \mathrm{otherwise}~.
                \end{cases} 
\end{aligned}
\end{equation}
The reference state for $\mathcal{H}_2$ is
\begin{equation}
\begin{aligned}
    n_{ie} &= \begin{cases} 
                1 & i = 1,6,11,13,21\\
                0 & \mathrm{otherwise}
                \end{cases} \\
    n_{i\mu} &=\begin{cases} 
                1 & i = 26,29,34\\
                0 & \mathrm{otherwise}~, 
                \end{cases} 
\end{aligned}
\end{equation}
and the reference state used for $\mathcal{H}_3$ is
\begin{equation}
\begin{aligned}
    n_{ie} &= \begin{cases} 
                1 & i = 1,6,9,11,13,21\\
                0 & \mathrm{otherwise}
                \end{cases} \\
    n_{i\mu} &=\begin{cases} 
                1 & i = 26,27,29,34\\
                0 & \mathrm{otherwise} ~.
                \end{cases} 
\end{aligned}
\end{equation}
The momentum modes are labeled in an increasing order of $z_y$ and $z_x$, e.g., 
\begin{align}
    {\bf z}_1 &= (1, -4) \;\text, ~~~ {\bf z}_6 = (3, -3) \;\text, ~~~ {\bf z}_9 = (2, -2)\nonumber\\
    {\bf z}_{11} &= (4, -2) \;\text, ~~~ {\bf z}_{13} = (2, -1) \;\text, ~~~ {\bf z}_{21} = (1, 1)\nonumber\\
    {\bf z}_{26} &= (2, 2) \;\text, ~~~ {\bf z}_{27} = (3, 2) \;\text, ~~~ {\bf z}_{29} = (1, 3)\nonumber\\
    {\bf z}_{34} &= (2, 4) \;\text. \label{eq:initp}
\end{align}
From these reference states, $H^{(K)}_{\nu\nu}$ populates a total of 14, 18, or 20 momentum modes for $\mathcal{H}_{1,2,3}$, thus allowing us to study the effects of non-forward scattering. The dimension of the Hilbert space becomes $d_h = 158, 1434$, and 6922 for $\mathcal{H}_{1,2,3}$ respectively.

\begin{center}
\begin{table}
\begin{tabular}{ | c || c | c | c | c | c | c|}
\hline
 Hilbert space & $~N_e~$ & $~N_{\mu}~$ & $~k~$  & Total  $~{\bf z}~$ &  $~~\sum_i |{\bf z}_i|~~$ & $~d_{h}~$\\ 
 \hline\hline
 $\mathcal{H}_1$ & 4 & 2 & 14 & (13,0)& 23.30 & 158\\
 \hline
 $\mathcal{H}_2$ & 5 & 3 &  18  & (16,0)& 26.95 & 1434\\
 \hline
 $\mathcal{H}_3$ & 6 & 4 & 20  & (21,0)& 33.38& 6922 \\
 \hline
\end{tabular}
\caption{Some defining quantities of the three Hilbert subspaces we study, that is, the number of neutrinos $N_e, N_{\mu}$, the number of momentum modes $k$ involved in the time evolution, total momentum, total kinetic energy, and the dimension of the Hilbert space ($d_h$). \label{table:h}}
\end{table} 
\end{center}

By construction, any initial state in $\mathcal{H}_{i}$ ($i=1,2,3$) stays in  $\mathcal{H}_{i}$ 
when evolved with  $H^{(K)}$.
For the sake of simplicity, we study the time evolution of 15 basis states for each case.
These initial states have the same momentum content as the reference states but have different flavor contents, with the total number of electron neutrinos  fixed to $N/2+1$. For $N=6$, there are 15 such basis states since $_6C_4 = 15$. For $N=8,10$ we picked 15 states from $_8C_5$ or $_{10}C_6$ number of such basis states. These initial states are far from equilibrium in both flavor and kinetic degrees of freedom. We leave the study of a more realistic initial state that mimics the situation in hot dense media of neutrinos to future work.

The time evolution of these initial states is performed by exactly diagonalizing the full or truncated Hamiltonian and applying the corresponding unitary time-evolution operator $e^{-i Ht}$ to the initial state. The Hamiltonian parameters are chosen to be $\sin 2\theta=0.8$ and $\bar\omega=0$. The diagonal terms in $H_{\mathrm{Kin}}$ proportional to $\bar T$ are dropped, since they are proportional to the 
identity in the restricted Hilbert spaces. For the rest of the section, we study various quantities that charachterize the time evolution 
of the chosen initial states: Loschmidt echo, $N_i^{+/-}$, and the one-body entropies introduced in  Sec.~\ref{sec:rhoands}.
 
We close this section by introducing a microcanonical ensemble 
for the systems we study~\cite{Deutsch:1991msp}. 
All basis states of a given Hilbert space, $\mathcal{H}_{i}$ ($i=1,2,3$), 
have the same kinetic energy, 
which dominates the energy of the basis states given the hierarchy Eq.~(\ref{eq:hierarchy}). 
Since all the basis states $| j \rangle$  in   $\mathcal{H}_{i}$
have the same particle numbers $N_e, N_{\mu}$, and total energy, 
in a microcanonical ensemble they are equally probable and the corresponding density operator is 
\begin{equation}
\label{eq:mc}
    \rho_{\mathrm{mc}} =  \frac{1}{d_h} \sum_{j=1}^{d_h} |j\rangle \langle j|~.
\end{equation}
In equilibrium, the expectation values of various observables $\mathcal{O}$ are computed as $\mathrm{Tr}\left[ \rho_\mathrm{mc} \mathcal{O} \right]$. 
The equilibrium expectation values of $N^+$ and $N^-$ will be  used to in later sections to quantitatively assess kinetic and flavor equilibration of our models. These equilibrium values are shown in the right panel of Fig.~\ref{fig:grid} for $\mathcal{H}_3$.


\subsection{Loschmidt echo}
\begin{figure}
    \centering
        \includegraphics[width=0.5\textwidth]{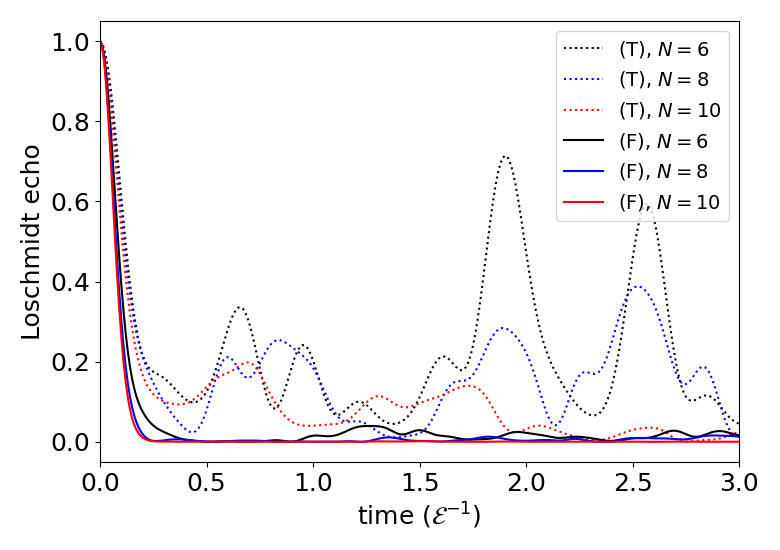}
        \caption{Loschmidt echo over time for the three cases,  $N=6$ (in $\mathcal{H}_1$), $N=8$ (in $\mathcal{H}_2$), and $N=10$ (in $\mathcal{H}_3$). Results from the time evolution under the full Hamiltonian, (F), are shown in solid lines, while truncated ones, (T), are in dotted lines. For each case, we chose a basis state in $\mathcal{H}_1$, $\mathcal{H}_2$, or $\mathcal{H}_3$ as the initial state. We turn off the vacuum oscillation ($\bar\omega=0$). \label{fig:LE_demo}}
\end{figure}

\begin{figure*}
    \centering
    \includegraphics[width=0.47\textwidth]{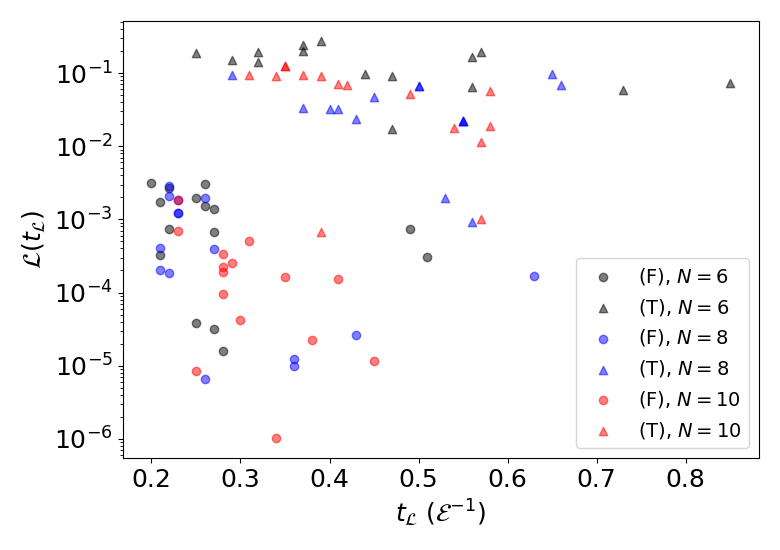} 
    \hspace{.5cm}
    \includegraphics[width=0.47\textwidth]{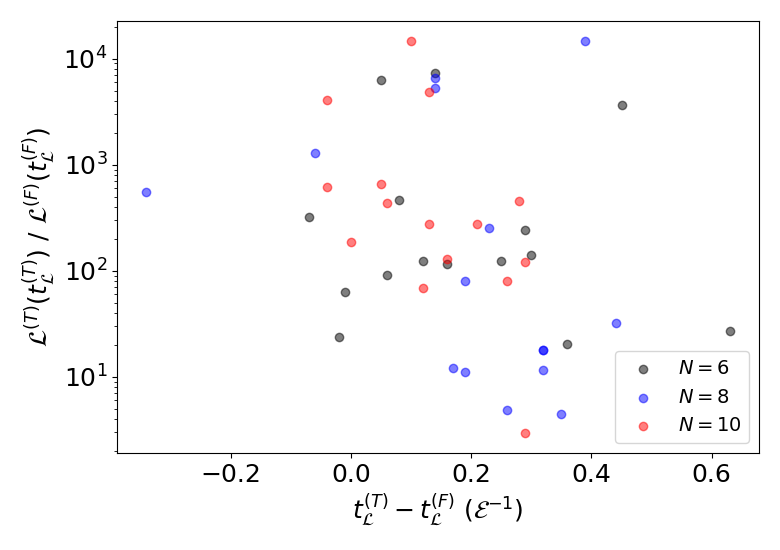} 
    \caption{
    Left: Loschmidt echo time scale $t_{\mathcal{L}}$ when the echo reaches its first minimum and the echo at the time. The initial states chosen in Sec.~\ref{subsec:2dkec} are evolved either via the full and truncated Hamiltonian. These initial states differ only by the flavor content of each momentum mode, but the populated momentum modes and total number of electron (and muon) neutrinos are fixed. Due to this choice, combined with the artifact of the momentum grid, some simulations yield the same Loschmidt echo. Right: the difference between $t_{\mathcal{L}}$ and ratio of the Loschmidt echo at $t_{\mathcal{L}}$ are shown for each initial state. \label{fig:LT}}
\end{figure*}

We begin the comparison of the time volution under the full and truncated Hamiltonian by studying the Loschmidt echo $\mathcal{L} \equiv |\langle \psi (t) | \psi(0) \rangle |^2$. At early time, the decrease of the Loschmidt echo indicates how quickly the amplitude of the wave function spreads from the initial state to the rest of the Hilbert space. The curvature of the echo at time $t=0$ is the negative of the variance of the Hamiltonian $\langle \psi(0) | H | \psi(0) \rangle ^2 - \langle \psi(0) | H^2 | \psi(0) \rangle$. This variance naively quantifies how densely the initial state is connected to the rest of the Hilbert space via the Hamiltonian,  since the $H^2$ piece measures the sum of  the square moduli of the Hamiltonian matrix elements between the initial state and all other states. In particular in our simulation setup, since the initial state is taken to be one of the basis states $|\psi(0) \rangle = |i\rangle$, 
the curvature of the Loschmidt echo at $t=0$ is 
\begin{equation}
    \frac{d^2}{dt^2} |\langle \psi (t) | \psi(0) \rangle |^2  \Big \vert_{t=0} = -  \sum_{j \neq i} |\langle i | H |j \rangle|^2\;\text.
\end{equation}
Therefore one expects that the Loschmidt echo decreases more rapidly with the full Hamiltonian than with the truncated Hamiltonian for a given initial state. 

Early-time behavior of the Loschmidt echo is shown for one of 15 initial states for $N=6,8,10$ in Fig.~\ref{fig:LE_demo}. As expected, the 
Loschmidt echo 
evolved under the full Hamiltonian (solid lines) decreases faster to lower values compared to those form the truncated simulation (dotted lines). Although the Loschmidt echo does not give separate information about the time scale of flavor evolution or kinetic randomization, it can still provide us with the evidence that the full Hamiltonian is able to spread the amplitude of the wave function more quickly to a larger Hilbert space than what the truncated Hamiltonian can achieve. Note that the Loschmidt echo from the full simulations especially with $N=8$ and 10 show rather quick convergence, whereas the others fluctuate about $\sim 0.3$. This is due to the smallness of the number of neutrinos and a resulting small dimension of the Hilbert space. It is demonstrated in \cite{Martin:2023gbo} that the Loschmidt echo decreases and converges quickly for $N=10 - 16$ systems under the truncated time evolution. We have confirmed this behavior in our simulations for $N=10 - 14$. 

To quantify the difference in randomization time scale under the full or truncated Hamiltonian, we introduce  $t_{\mathcal{L}}$, 
defined as the time at which the Loschmidt echo reaches its first minimum. In Fig.~\ref{fig:LT}, we show $t_{\mathcal{L}}$ and the Loscmidt echo at the time for all 15 initial states in all $N$ on the left panel. The circles show $t_{\mathcal{L}}$ obtained in the wave function evolved by the full Hamiltonian, while triangles correspond to the truncated Hamiltonian. The difference between $t_{\mathcal{L}}$ and the Loschmidt echo at the time $t_{\mathcal{L}}$ by $H^{(F)}$ and $H^{(T)}$ is striking. The full time evolution, on average, achieves a minimum Loschmidt echo a few order of magnitude smaller, in a shorter time. To emphasize this point, we compare the time scale for each initial state from the full and truncated Hamiltonian on the right panel of Fig.~\ref{fig:LT}. The figure shows $t^{(T)}_{\mathcal{L}} - t^{(F)}_{\mathcal{L}}$, i.e., the difference of $t_{\mathcal{L}}$ between two Hamiltonians on the horizontal axis and the ratio of the echos $\mathcal{L}^{(T)}/\mathcal{L}^{(F)}$ on the vertical axis. For almost all initial states taken for $N=6,8,10$, $t_{\mathcal{L}}$ is smaller with the full time evolution and Loschmidt echo is smaller at the first minimum. Within our study, we are not able to observe an obvious trend in $t_{\mathcal{L}}$ as we vary the number of neutrinos $N$. We will leave the study of the behavior of the Loschmidt echo in systems with larger $N$ to future work.


\subsection{Single neutrino observables}
\label{sect:sno}

\begin{figure*}
    \centering
    \includegraphics[width=0.47\textwidth]{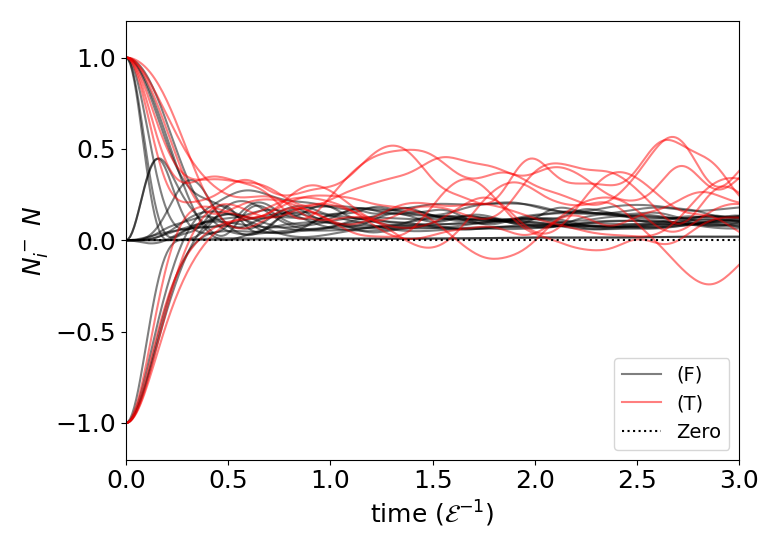} 
    \hspace{.5cm}
    \includegraphics[width=0.47\textwidth]{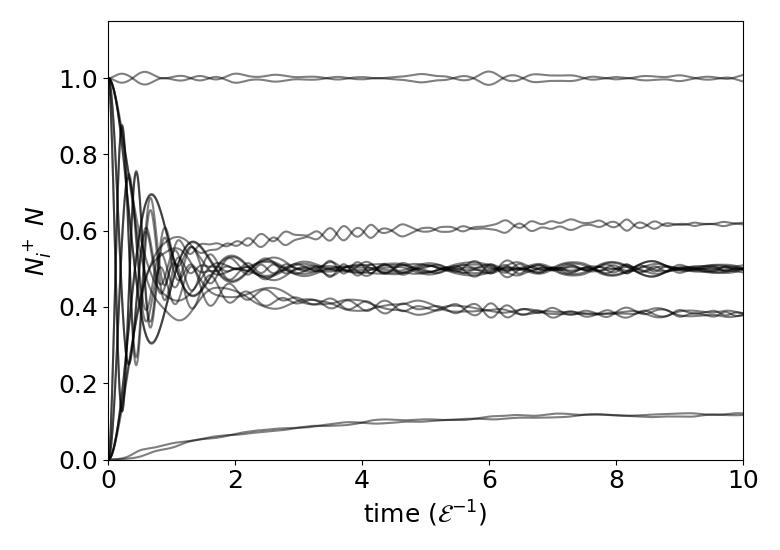} 
    \caption{ 
    Left: $N^{-}_{i}$ (normalized to 1) for each momentum mode ${\bf p}_i$ over time. The initial state is a basis state in  $\mathcal{H}_3$ 
    (with $N=10$) and is time evolved under the full (F) or truncated (T) Hamiltonian. Right: $N^+_i$ (normalized to 1) for each momentum mode ${\bf p}_i$ over time from the same simulation as the left panel. \label{fig:Nn10_npm}}
\end{figure*}

In this section we analyse kinetic randomization and flavor evolution of our neutrino systems separately through the occupation numbers $N^+, N^-$ introduced in Eq.~(\ref{eq:Nipm}). In particular, we show that these expectation values converge to their equilibrium values from the microcanonical ensemble in Eq.~(\ref{eq:mc}), thus demonstrating the equilibration process of our models. 

The neutrino flavor degrees of freedom evolve over time due to their vacuum oscillation (which we do not consider here by setting $\bar \omega =0$) 
and self-interaction. While forward scattering can cause flavor equilibration by exchanging flavors among neutrinos~\cite{Martin:2023gbo}, non-forward scattering processes are expected to speed-up the equilibration process by activating the modes which are inaccessible via forward scattering and letting flavor mixing happen within those modes. To visualize this effect, we inspect $N_i^-$, which is the  
difference between  the occupation number of flavors $e$ and $\mu$ for the momentum mode ${\bf z}_i$. As an example,  
in the left panel of Fig.~\ref{fig:Nn10_npm} we show $N^-_i$ over time for all 
modes ${\bf z}_i$ from the time evolution of a basis state in $\mathcal{H}_3$ ($N=10$). 
When using the truncated Hamiltonian, only the 10 initially populated bins evolve (shown in red), 
starting at either $N \, N_i^-=1$ (electron neutrino) or $N  \, N_i^-=-1$ (muon neutrino). 
On the other hand, when using the full Hamiltonian, 
10 additional momentum modes are activated via non-forward scattering processes through the time evolution. 
We see this effect in Fig.~\ref{fig:Nn10_npm} already at early time $t\sim 0.2/\mathcal{E}$. A closer look at the evolution of $N_i^-$ at very early time indicates an acceleration of flavor evolution via non-forward scatterings. 
We will quantify this acceleration in more detail in the next section by looking at the one-body entropy. 

\begin{figure}
    \centering
    \includegraphics[width=0.47\textwidth]{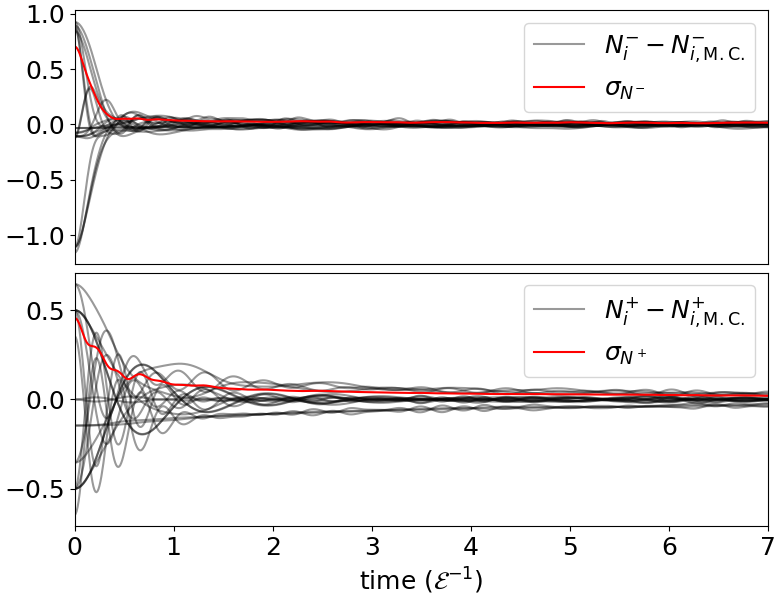}
    \caption{\label{fig:npm-MC} Top:  deviation of $N^{-}_i$ from the equilibrium values
        (computed with Eq.~(\ref{eq:mc})) for all momentum modes, along with the standard deviation ($\sigma_{N^-}$). 
  Bottom:  deviation of $N^{+}_i$ from the equilibrium values for all momentum modes, along with the
   standard deviation  ($\sigma_{N^+}$). The initial state is the same as the one shown in Fig.~\ref{fig:Nn10_npm}.}
\end{figure}

After $N^{-}_i$s decrease to nearly zero for both full and truncated evolutions around $t\sim 0.5/\mathcal{E}$, $N^-_i$s show narrower fluctuations with the full Hamiltonian than with the truncated Hamiltonian. In the full evolution, $N_i^{-}$ (normalized to 1) converges to the equilibrium values (right panel of Fig.~\ref{fig:grid}) quickly and fluctuate around the equilibrium by $\lesssim 0.02$. This is shown in the upper panel in Fig.~\ref{fig:npm-MC} --- the black lines are the difference of $N^{-}_i$ from the microcanonical value, $N^{-}_i - N^{-}_{i, \mathrm{M.C.}}$, for all 20 momentum modes and the red line ($\sigma_{N^-}$) shows the standard deviation of the difference across the 20 modes. Clearly $N^{-}_i$s converge to the equilibrium values at around $t = 0.5/\mathcal{E}$ and the fluctuate around equilibrium. This is the first evidence we show regarding the equilibration of flavor degrees of freedom in our models. We will further confirm this finding in the next section.

Kinetic randomization of many-body neutrino systems is what we observe only in the presence of the non-forward self-interactions. On the right panel of Fig.~\ref{fig:Nn10_npm}, we show $N N_i^{+}$ for all momentum modes for a 10-neutrino simulation in the Hilbert space $\mathcal{H}_3$ --- the same initial state as the one shown on the left panel for $N^-$ is used. Our results indicate  that momentum mode occupation numbers 
start fluctuating around  asymptotic values for times $t \sim (2-5)/\mathcal{E}$, depending on the momentum mode. 
The convergence of $N^+$ to equilibrium is shown in the lower panel of in Fig.~\ref{fig:npm-MC} --- the black lines are the difference of $N^{+}_i$ from the microcanonical value for all 20 momentum modes and the red line ($\sigma_{N^+}$) shows the standard deviation of the difference across the 20 modes. 
The deviation of $N_i^{+}$ from equilibrium does not appear to drop as quickly as the deviation of $N_i^-$ from equilibrium. 
This could be attributed to the fact that 
it takes time for some momentum modes to be populated from a certain initial state due to the smallness of the $g$ factors -- 
an effect that we expect to disappear in larger systems and for initial states with more isotropic momentum distribution.

To summarise, the time evolution of $N_i^+$ and $N_i^-$ in our model with $N=10$ neutrinos shown in Fig.~\ref{fig:Nn10_npm} demonstrate two things. First, non-forward scattering change the flavor evolution of the state by activating momentum modes that are initially unoccupied. It appears that the activation of unoccupied modes accelerate the flavor equilibration. Second, even in such a small system of neutrinos, the initial state that is far from equilibrium can kinetically thermalize. The time scales for flavor and kinatic equilibration 
appear to be both $t\sim O(1)/\mathcal{E}$, 
and we cannot identify any dependence on the total number of neutrinos $N$.
Further studies will be 
needed to 
explore whether a separation of the two time scales 
arise.  Kinetic theory suggests that the time scale for momentum evolution 
should scale as  $t_{\rm inc} \sim (G_F^2 T^5)^{-1}$, in accordance with Fermi's golden rule,  
while the time scale for coherent refractive effects should scale as   $t_{\rm coh} \sim (G_F T^3)^{-1}$. 
These estimate rely on the assumption (built into kinetic theory) of  interactions localized in 
space and time, which on the many-body side would require working with wave-packets, which can be built 
out of our plane waves basis. 
If one works with  plane waves in a box, as shown in Ref.~\cite{Friedland:2003eh} 
the coherent and incoherent time scales become 
$t_{\rm coh} \sim 1/({\cal E} N)$  and $t_{\rm inc} \sim 1/({\cal E} \sqrt{N})$, respectively. 
Therefore, we expect that to disentangle  the two will require studying larger systems, with $N \sim O(100)$.

\subsection{Entanglement entropy}

\begin{figure}
    \centering
        \includegraphics[width=0.5\textwidth]{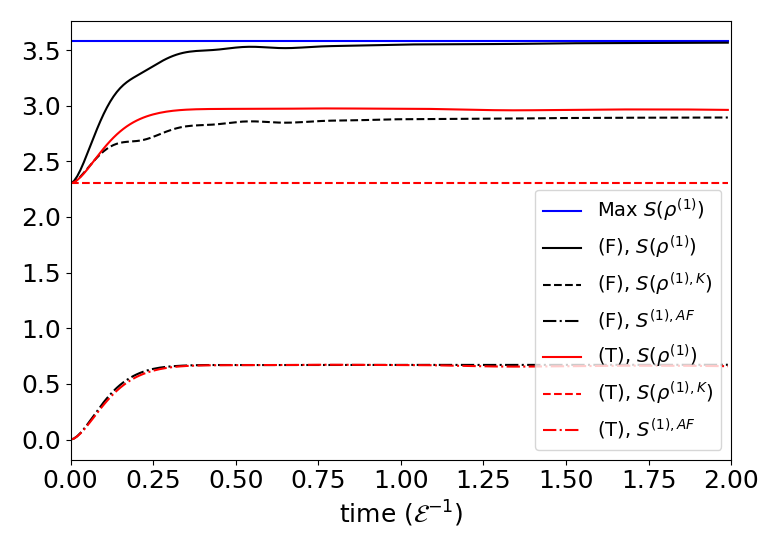}
        \caption{ \label{fig:S_comp}  Time evolution of the one-body entropy $S(\rho^{(1)})$, its kinetic component $S(\rho^{(1),K})$, and flavor component $ S^{(1),AF}=S(\rho^{(1)}) - S (\rho^{(1),K})$ for a 10-neutrino basis state. The black lines show the entropies 
        from  the time evolution with the full Hamiltonian, while the red lines show the entropies  from the time evolution
        generated by the truncated Hamiltonian. The maximal entropy computed from the microcanonical ensemble is shown by the blue line.}
\end{figure}

Finally, we quantitatively assess the time scales of the flavor evolution and kinetic thermalization via the one-body  entropies $S(\rho^{(1)}), S(\bar\rho^{(i)})$ and $S(\rho^{(1,K)})$, which are introduced in Eq.~(\ref{eq:ee1}) and Eq.~(\ref{eq:ek1}) respectively. The one-body entropy $S(\rho^{(1)})$ quantifies the entanglement of one neutrino with the rest of the system, and it can be divided to two parts: the kinetic entanglement $S(\rho^{(1,K)})$ and flavor entanglement $(S^{(1),AF}=S(\rho^{(1)})-S(\rho^{(1,K)})$. In Fig.~\ref{fig:S_comp}, we show these one-body entropies of a 10-neutrino state over time. The initial state is the same $N=10$ basis state as the one used in Fig.~\ref{fig:Nn10_npm}. The black lines show the entropies from the full simulation, while the red lines show those from the truncated evolution. For each Hamiltonian, $S(\rho^{(1)}), S(\rho^{(1,K)})$, and $S^{(1), AF}$  are shown in solid line, dashed line, and dash-dotted line respectively. At initialization, the flavor component of the entropy, $S^{(1), AF}$, is zero since each neutrino has definite flavor $e$ or $\mu$. On the other hand, since 10 momentum modes are occupied, the kinetic component, $S(\rho^{(1,K)})$, starts at a non-zero value. Under the time evolution with the full Hamiltonian, both the kinetic and flavor component of the entropy grow and their sum asymptotes to the maximal value predicted via the microcanonical ensemble (shown in the blue line). Under the truncated evolution, while the flavor component of the entropy grows over time, the kinetic component stays constant due to the lack of non-forward scattering processes. In this example, the flavor component of the entropy grows at very similar rate with the full or truncated evolution. We will take a closer look at the difference between flavor evolution under the two Hamiltonians later in the section.

\begin{figure}
    \centering
        \includegraphics[width=0.5\textwidth]{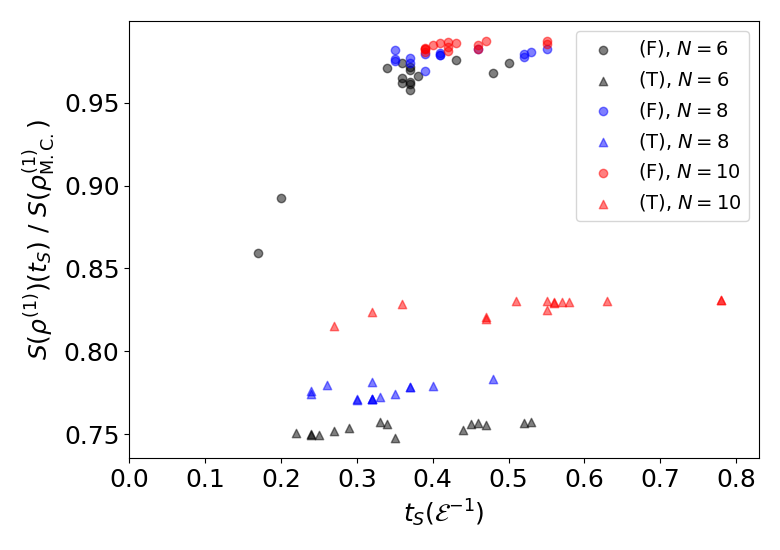}
        \caption{\label{fig:s1_all} Time at which the one-body entropy $S(\rho^{(1)})$ reaches its first maximum, $t_S$, and the entropy at $t_S$ is shown for 15 initial states with $N=6$ (black), $N=8$ (blue), or $N=10$ (red). Circles show the results with 
        obtained with the full Hamiltonian,  
        while the triangles show those obtained with the   truncated Hamiltonian.}
\end{figure}

To quantify the time scale associated to the growth of  entropy for $N=6,8,10$, 
we define $t_S$ to be the time when the one-body entropy $S(\rho^{(1)})$ reaches its first maximum. We show $t_S$ and the value of entropy at $t_S$ divided by the maximal value via microcanonical ensemble in Fig.~\ref{fig:s1_all} for $N=6,8,10$ neutrinos evolved from the 15 initial states introduced in Sec.~\ref{subsec:2dkec}. Most notably, for the full simulation shown in circles, most initial states reach over $95\%$ of the maximal entropy around $t_S\sim0.5 /\mathcal{E}$. Interestingly, among the systems we study, we do not see much difference in $t_S$ as the number of neutrino is varied. Note that the initial states will not be able to reach the maximal entropy under the truncated time evolution, as shown by the triangles in Fig.~\ref{fig:s1_all}. This is because the kinetic component of the entropy can not be maximized due to the lack of non-forward scattering processes.

To gain insight on the flavor and kinetic equilibration time scales,  we 
look into the kinetic and flavor component of the entropy separately. 
Regarding the flavor evolution,  the flavor component of the one-body entropy is a weighted (by $N^+_i/N$) sum of the one-mode entropy $S(\bar\rho^{(i)})$ as is shown in Eq.~(\ref{eq:ee1}). Therefore, to remove the 
information on the kinetic randomization, 
we will first study the one-mode entropy directly.
Following that, we will study the time scale of kinetic randomization from the kinetic component $S(\rho^{(1,K)})$ 
of the one-body entropy. 

\begin{figure}
    \centering
        \includegraphics[width=0.5\textwidth]{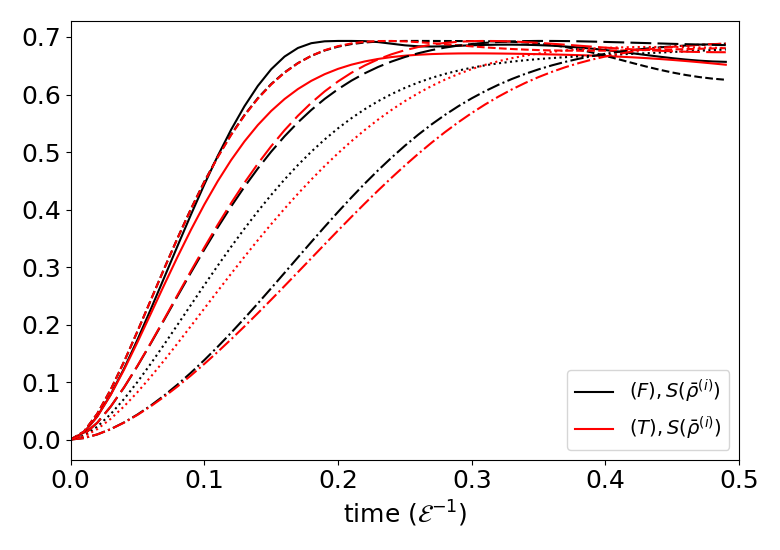}
        \caption{\label{fig:Sbar_N10}   The entropy per momentum bin, $S(\bar\rho^{i})$, is shown for 5 (${\bf z}_1,{\bf z}_6,{\bf z}_9,{\bf z}_{11},{\bf z}_{13}$) out of 10 momentum modes that are initially occupied in a 10-neutrino basis state. Black lines show the entropy 
        computed via the full Hamiltonian, while the red lines show those from the truncated Hamiltonian. The line patterns characterizes 
        a given momentum bin, regardless of the Hamiltonian used for the time evolution.}
\end{figure}

\begin{figure*}
    \centering
        \includegraphics[width=1.0\textwidth]{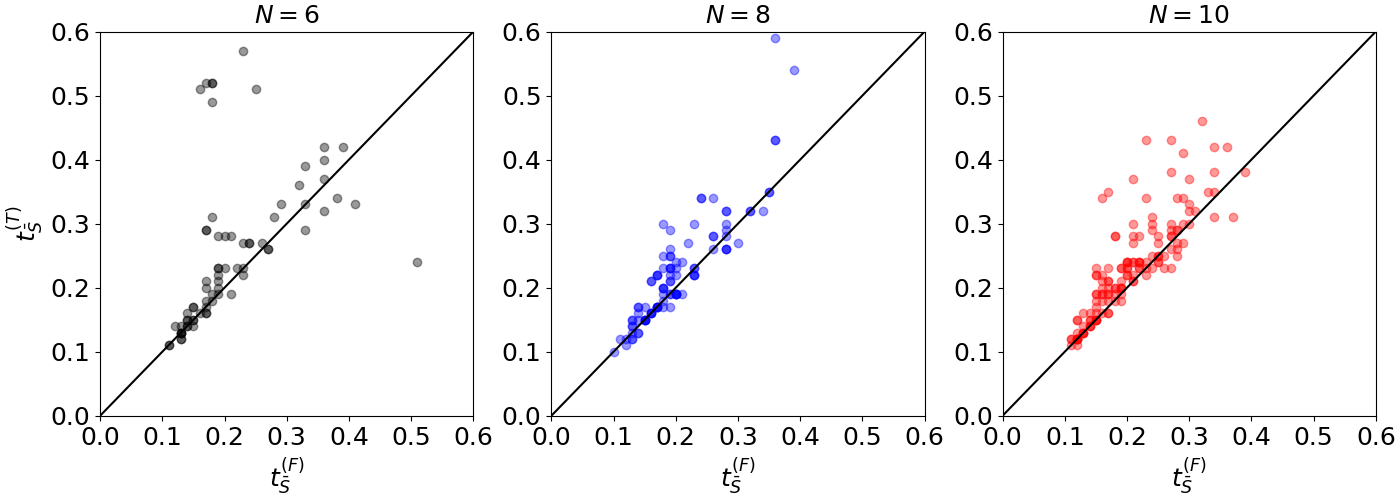}
        \caption{\label{fig:Sbar_90_linear}  The time scale $t_{\bar S}$ from the full and truncated evolution are on the horizontal and vertical axis respectively. Results from all 15 simulations and for all momentum modes in Eq.~(\ref{eq:initp}) are shown together. The solid lines show the line of $t^{(F)}_{\bar S} = t^{(T)}_{\bar S}.$
        }
\end{figure*}

\begin{figure*}
    \centering
        \includegraphics[width=1.0\textwidth]{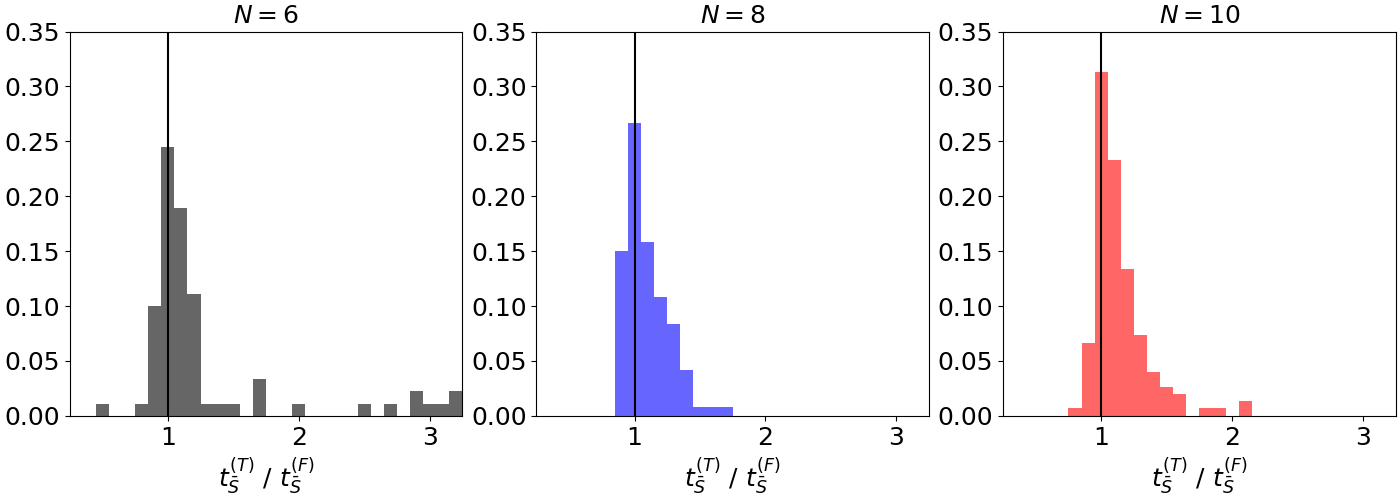}
        \caption{\label{fig:Sbar_90_ratio} The histogram of the ratio $t^{(T)}_{\bar S} / t^{(F)}_{\bar S}$ (truncated/full) in the range $<3.25$. Results from all 15 simulations and for all momentum modes in Eq.~(\ref{eq:initp}) are counted in together. The histogram is normalized by the total number of counts, that is 90, 120, 150 for $N=6,8,10$ respectively.
        }
\end{figure*}

The entropy of the one-mode (normalized) density matrix $\bar\rho^{(i)}$ quantifies the 
entanglement of  neutrinos 
due to their 
flavor degrees of freedom. As was hinted via the 2-neutrino toy model in Fig.~\ref{fig:toy_bin1}, flavor evolution can be accelerated by non-forward scattering processes since they open up momentum modes that are initially unoccupied, and flavor evolution can take place within those new sectors of the Hilbert space. To illustrate this observation, in Fig.~\ref{fig:Sbar_N10}, we show the one-mode entropy $S(\bar\rho^{(i)})$ over time for 5 (${\bf z}_1,{\bf z}_6,{\bf z}_9,{\bf z}_{11},{\bf z}_{13}$) out of 10 momentum modes that are occupied in the $10$-neutrino initial state. The entropy of the wave function evolved under the full and truncated Hamiltonians are shown in black and red lines respectively. The same line type is used for the entropy of the same momentum mode from the full and truncated time evolution. The figure shows that the entropy grows slightly faster under the full time evolution than via the truncated one (except those in short dashed lines which behave nearly the same for full and truncated evolution).

To quantify the time scale of flavor evolution, we introduce $t_{\bar S}$ to be the time when the one-mode entropy reaches $90\%$  of the maximal value, i.e., $0.9\times\log 2$, for the first time. In Fig.~\ref{fig:Sbar_90_linear}, we show the time scales $t^{(F/T)}_{\bar S}$ (full or truncated evolution) on the horizontal and vertical axis respectively for initial momentum modes in all 15 simulations. 
 Since we focus on the momentum modes in Eq.~(\ref{eq:initp}) that are occupied at $t=0$ and do not count in those that are populated later by the non-forward scattering processes, the total number of data points should be 90 for $N=6$, 120 for $N=80$, and 150 for $N=10$ for both full and truncated set-ups. However, we discarded those $t_{\bar S} > 0.6 /\mathcal{E}$, thus 17 ($N=6$), 20 ($N=8$), or 9 ($N=10$) data points from the truncated simulations are excluded from the plots. 
 The distribution of points around  
 the line of $t^{(F)}_{\bar S} = t^{(T)}_{\bar S}$ in black demonstrates the speed-up of flavor evolution via non-forward scattering processes even in these small systems. To show this result even more explicitly, 
in Fig.~\ref{fig:Sbar_90_ratio}
 we plot the normalized histogram of the ratio $t^{(T)}_{\bar S}/t^{(F)}_{\bar S}$ for the range $\left[0,3.25\right]$. The bin with range $\left[0.95,1.05\right)$ is marked with the black vertical line. By comparing the bins on the left and right side of the vertical line, one can clearly see 
 that
 the non-forward processes 
 affect the time scale of flavor evolution for individual momentum modes 
 and 
 induce a bias towards faster equilibration. 
 
 While we observed acceleration of flavor evolution via the full Hamiltonian in our small models, it is difficult to conclude much about the dependence of the time scales on $N$ and $k$. We will leave a study of flavor evolution time scale with larger number of neutrinos and/or larger number of momentum modes to future work, where we expect to see further speed-up in the equilibration in the flavor degrees of freedom. 

\begin{figure}
    \centering
        \includegraphics[width=0.5\textwidth]{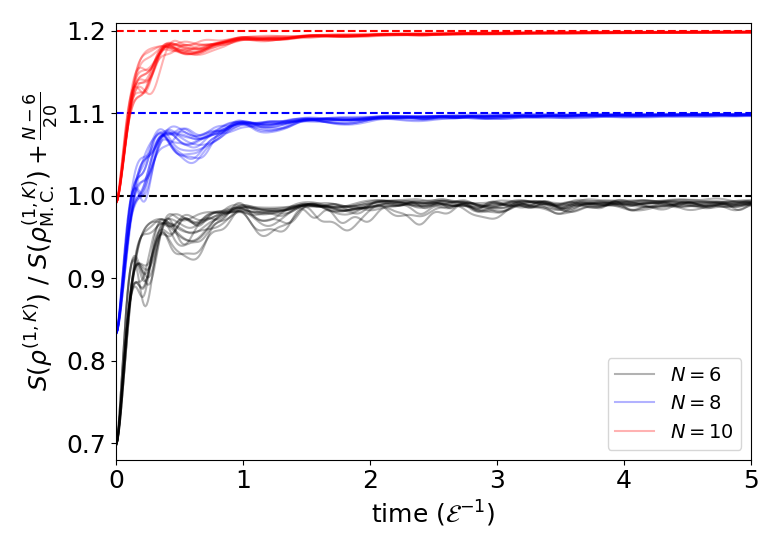}
        \caption{The ratio of the kinetic component of the one-body entropy, $S^{(1,K)}$, to its maximal value in equilibrium is shown over time for   15 initial states. The lines $N=8$ (blue) and $N=10$ (red) are shifted by 0.1 or 0.2 for better visualization. The dotted lines show the maximal entropy predicted from the microcanonical ensemble. \label{fig:Sk_90}}
\end{figure}

We close the section by demonstrating kinetic thermalization in our neutrino systems 
evolved with the full Hamiltonian. The  kinetic properties of many-body neutrino systems are captured by the occupation numbers $N^+_i$, and their expected values in equilibrium can be computed from the microcanonical ensemble in Eq.~(\ref{eq:mc}). Equivalently, the kinetic one-body entropy $S^{(1,K)}$ should increase over time and asymptote to the maximal value predicted from the microcanonical ensemble as the system equilibrates. 
In Fig.~\ref{fig:Sk_90}, we show the ratio of $S^{(1,K)}$ to its maximal value in microcanonical ensemble over time for the 15 initial states in black, blue, and red lines for $N=6,8,10$ cases respectively. The ratios for $N=8, 10$ are shifted by 0.1 and 0.2, respectively. The values of  $S^{(1,K)}$ in equilibrium are 2.567 for $N=6$, 2.837 for $N=8$, and 2.909 for $N=10$. The systems with 6 neutrinos have trouble completely thermalizing in our model. On the other hand, we see a nice convergence of the entropy to the maximal value for 8 and 10-neutrino states. 

In this section, we analysed the one-body entropy and the time scales $t_S$ and $t_{\bar S}$, that are defined according to the growth of entropy over time. Under the full time evolution, the one-body entropy $S(\rho^{(1)})$ reaches  the first maximum for 6, 8, and 10-neutrino systems around $t\sim0.5 /\mathcal{E}$, and we do not see an obvious trend as we vary the number of neutrinos. 
According to the definition of $S(\rho^{(1)})$, the maximal entropy 
cannot be achieved using the truncated Hamiltonian
due to the lack of non-forward scattering processes. Flavor evolution is where we can directly compare  the full and truncated Hamiltonian. First, we show that flavor equilibration demonstrated in Ref.~\cite{Martin:2023gbo} in the forward limit occurs with the full Hamiltonian as well. Furthermore, we show  evidence within our models that the full Hamiltonian changes the flavor evolution in the direction of accelerating the equilibration. 
Thermalization of the momentum distribution is demonstrated in our model, and the time scale for kinetic thermalization 
appears to be slightly longer longer than the flavor equilibration time scale. 
We do not have  evidence that these time scales are independent of the models we chose. 
In particular, the separation of two time scales can vary with the number of neutrinos or 
available momentum modes. We leave the study of these time scales 
with larger number of neutrinos and/or momentum modes to future work.

\section{Conclusions and outlook} 
\label{sect:discussion}

In this paper we studied neutrino flavor evolution in the quantum many-body approach  using 
the full neutrino-neutrino Hamiltonian,  
going beyond the commonly adopted truncated version that allows only for 
the couplings of neutrino pairs satisfying the forward kinematics condition.

We have set up a framework to implement the time evolution 
of the system using the occupation number representation for the many-body system. 
In this setup, we have explored  the evolution of  simple initial states, 
i.e.  product states of neutrinos with different flavor and momenta. 
For simplicity, we have restricted the analysis to two flavors. 
We have studied a toy model with $N=2$ neutrinos and models 
with  momenta on a two-dimensional  grid with $N=6,8,10$ neutrinos 
and up to $k=20$ 
momentum modes.
We have quantified the time scales for 
evolution of flavor and momentum degrees of freedom, 
and their interplay. 
The main lessons from our explorations can be summarized as follows: 
\begin{itemize}

\item  
The hierarchy between the neutrino kinetic and potential energy ($T \gg  G_F T^3$) 
results in dynamical pair-wise kinetic energy conservation. 
This is the statement that non-forward terms in $H_{\nu \nu}$, 
that couple incoming and outgoing pairs of neutrions, 
significantly affect the time evolution only when 
the difference in kinetic energy of the two pairs is on the order of or smaller than 
potential energy due to the self-interactions. 
This observation leads to simplifications in the algorithm for time evolution. 
Moreover, it is worth mentioning that 
together with three-momentum conservation, for each pair of momenta ${\bf p}$ and ${\bf q}$ 
this dynamical kinetic energy conservation 
still opens up an infinite set of momentum pairs ${\bf p}^\prime$ and ${\bf q}^\prime$   
that contribute to the evolution (see  Appendix~\ref{sect:appendix2}).     
A byproduct of this analysis is that one can only see significant differences between full 
and truncated evolution in systems with spacetime dimension $d>2$.

\item   On the qualitative side,  we find that non-forward processes 
affect the dynamics significantly. 
First,  even for the small systems with up to $N=10$ considered in this study, 
we find that  non-forward processes induce kinetic (momentum) randomization  
on top of the flavor randomization already induced by the truncated Hamiltonian~\cite{Martin:2023gbo}.
We observe `thermalization', i.e. convergence towards expectation values in 
a suitably defined  microcanonical ensemble, in both 
flavor and momentum on comparable time scales. 
We also observe that the inclusion of 
non-forward processes generates a faster flavor evolution compared to the one induced
 by the truncated (forward) Hamiltonian.

\item  On the quantitative side, we studied the impact on the evolution time scales 
using a number of  metrics, such as   the Loschmidt echo and the 
entanglement entropy associated with the one-body density matrix. 
The time scales in all observables are comparable, with  $t \sim O(1) /\mathcal{E}$.
    
\end{itemize}

In the many-body approach studied in this paper, several
open questions remain before one can draw definite conclusions about problems of astrophysical interest, 
such as  assessing the impact of neutrino flavor evolution 
on nucleosynthesis~\cite{Balantekin:2023ayx} and on the neutrino signal from galactic supernovae.
We can identify  several interesting thrusts for future investigations: 
(i) A key step 
is the study of systems with larger number of neutrinos ($N$), 
which  will enable a number of interesting investigations. 
These include studying the scaling of various observables and 
time scales with $N$;  exploring the possible emergence of coherent enhancements, 
e.g. by considering initial states with multiple neutrinos within a given solid angle; 
exploring the effect of spatially non-homogeneous initial conditions (neutrino 
wavepackets). 
These studies, besides their intrinsic interest, 
will also help clarifying the connection between the many-body approach and kinetic theory.
(ii) Include neutrino-matter interactions in our formalism. 
This requires the implementation of additional four-fermion operators in the Hamiltonian,  
which are however technically simpler than the one studied here because they 
are linear or bilinear in the neutrino field.
(iii) Implementation on a quantum computer, which requires 
finding efficient mappings of the full $H_{\nu \nu}$ onto qubit Hamiltonians.
(iv) Work towards a comparison with the QKEs, using as common ground the one-body density matrix. 
To quantitatively explore the connection of many-body and QKE approaches 
will require  simulating systems with larger number of neutrinos (see point (i) above and discussion at the end of Section~\ref{sect:sno}).

\begin{acknowledgements}
We acknowledge stimulating discussions with 
Baha Balantekin,
Joe Carlson, 
Huaiyu Duan, 
Alex Friedland, 
Julien Froustey, 
George Fuller, 
Luke Johns, 
Scott Lawrence, 
Gail McLaughlin, 
Josh Martin, 
Duff Neill, 
Amol Patwardhan,
Ermal Rrapaj, 
Sanjay Reddy, 
Alessandro Roggero, 
Martin Savage, and 
Irene Tamborra. 
We are very grateful to Julien Froustey and Ermal Rrapaj 
for cross-checking the form of our $H_{\nu \nu}$ with their unpublished work. 
We also thank Julien Froustey, Luke Johns, Scott Lawrence, Josh Martin, and
Duff Neill for providing comments on the manuscript.
V.C. and Y.Y. acknowledge support by the U.S. DOE Office of Nuclear Physics under Grant No. DE-FG02-00ER41132. SS acknowledges support from the Department of Energy, Nuclear Physics Quantum Horizons program through the Early Career Award DE-SC0021892.
\end{acknowledgements}

\newpage

\appendix

\section{Kinematics of $2 \to 2$ scattering}
\label{sect:appendix2}
Consider the reaction 
$\nu_\alpha ({\bf p}) + \nu_\beta ({\bf q})  \to \nu_{\alpha^\prime} ({\bf p}^\prime) + \nu_{\beta^\prime}  ({\bf q}^\prime)$.
We want to parameterize all the pairs of 3-vectors ${\bf p}^\prime$, ${\bf q}^\prime$ such that 
${\bf p} + {\bf q} = {\bf p}^\prime + {\bf q}^\prime$  and 
$|{\bf p}| + |{\bf q}| = |{\bf p}^\prime| + |{\bf q}^\prime|$ 
in terms of ${\bf p}$, ${\bf q}$, and two angles, denoted below by $\theta$ and $\phi$.
This is achieved by 
(i) boosting to the center of mass (CMS) system of the initial momentum pair 
$({\bf p}, {\bf q}) \to ({\bf p}_{CMS}, {\bf q}_{CMS})$; 
(ii) parameterizing the outgoing  momenta for elastic scattering  
$({\bf p}_{CMS}, {\bf q}_{CMS}) \to ({\bf p}^\prime_{CMS}, {\bf q}^\prime_{CMS})$ 
in terms of the polar and azimuthal angles $(\theta, \phi)$ of the 
unit vector $\hat{v} \equiv  {\bf p}^\prime_{CMS}   / |{\bf p}^\prime_{CMS}|$;  
(iii) boosting back to the original reference frame 
$({\bf p}^\prime_{CMS}, {\bf q}^\prime_{CMS}) \to ({\bf p}^\prime, {\bf q}^\prime)$.

Explicitly, in terms  of the CMS velocity 
$\boldsymbol{\beta} = ({\bf p} + {\bf q})/ (|{\bf p}| + |{\bf q}|)$, 
$\gamma = 1/\sqrt{1 - \beta^2}$,   the unit vector $\hat v = (\sin \theta \cos \phi, \sin \theta \sin \phi, \cos \theta)$, 
and $|{\bf p}_{CMS}|= |{\bf q}_{CMS}|= \gamma (|{\bf p}|  -  \boldsymbol\beta \cdot {\bf p})$, one has
\begin{eqnarray}
    {\bf p}^\prime &=& |{\bf p}_{CMS}| 
    \left( \hat v  + \boldsymbol{\beta}     \left( \gamma + \frac{\gamma -1}{\beta^2} \boldsymbol\beta \cdot \hat v \right) \right)
\\
{\bf q}^\prime &=& 
|{\bf q}_{CMS}| 
    \left( - \hat v  + \boldsymbol{\beta}     \left( \gamma - \frac{\gamma -1}{\beta^2} \boldsymbol\beta \cdot \hat v \right) \right)~.
\end{eqnarray}

\section{Hamiltonian including anti-neutrinos}
\label{sect:app_nubar}

In Section~\ref{sect:formalism} we explicitly wrote only  
the neutrino-neutrino part of $H_{\nu \nu}$, ignoring  anti-neutrinos. Here we will write down  
all terms in the   many-body Hamiltonian, taking into account both neutrinos and anti-neutrinos. 
Treating neutrino masses as a perturbation, we only include positive helicity anti-neutrino 
modes and use the notation $b_\alpha ({\bf p}, +) \to b_\alpha ({\bf p})$, with $\alpha = e, \mu$.

The kinetic part of the Hamiltonian for the anti-neutrinos can be obtained by  just replacing  
the $a_{e/\mu}$ by $b_{e/\mu}$ in Eq.~(\ref{kin}).
When considering the interaction terms,  
the $\bar \nu$-$\bar \nu$ and  $\nu$-$\bar \nu$ terms in   $H_{\nu \nu}$
are proportional to the same angular function  $g ({\bf p^\prime}, {\bf p}, {\bf q^\prime}, {\bf q} )$ 
that controls the $\nu$-$\nu$ interactions.
Keeping only the terms that conserve total particle number (i.e. discarding terms that mediate $\nu \leftrightarrow \nu \bar \nu \nu$) we find
\begin{widetext}

\begin{eqnarray}
    H_{\nu \nu} &=  & \frac{G_F}{\sqrt{2}} \ 
    \sum_{\alpha, \alpha^\prime, \beta, \beta^\prime} \ 
         \int  \frac{d{\bf q}}{(2 \pi)^3}  \frac{d{\bf q^\prime}}{(2 \pi)^3}  \frac{d{\bf p} }{(2 \pi)^3} \frac{ d{\bf p^\prime}}{(2 \pi)^3} 
        \  g ({\bf p^\prime}, {\bf p}, {\bf q^\prime}, {\bf q} )  \ \frac{\delta_{\alpha^\prime \alpha} \delta_{\beta^\prime \beta} + 
        \delta_{\alpha^\prime \beta} \delta_{\beta^\prime \alpha}}{2}
\       
\nonumber    \\ 
    & \times &  \  \Big(  a^\dagger_{\alpha^\prime} ({\bf p^\prime}) \, a_\alpha ({\bf p}) \, 
     a^\dagger_{\beta^\prime} ({\bf q^\prime}) \, a_\beta ({\bf q}) 
\ (2 \pi)^3 \delta ( {\bf p} + {\bf q}     - {\bf p^\prime} - {\bf q^\prime} )\nonumber\\
& + &  2 \  a^\dagger_{\alpha^\prime} ({\bf p^\prime}) \, a_\alpha ({\bf p}) \, 
     b_{\beta^\prime} ({\bf q^\prime}) \, b^{\dagger}_\beta ({\bf q}) 
\ (2 \pi)^3 \delta ( {\bf p} - {\bf q}     - {\bf p^\prime} + {\bf q^\prime} )
\  
      \  
      \nonumber\\
& + & 2  \  b_{\alpha^\prime} ({\bf p^\prime}) \, b^{\dagger}_\alpha ({\bf p}) \, 
     a^\dagger_{\beta^\prime} ({\bf q^\prime}) \, a_\beta ({\bf q})   
\ (2 \pi)^3 \delta ( {\bf p} - {\bf q}     - {\bf p^\prime} + {\bf q^\prime} )
\  
      \nonumber\\
& + &  \  b_{\alpha^\prime} ({\bf p^\prime}) \, b^{\dagger}_\alpha ({\bf p}) \, 
     b_{\beta^\prime} ({\bf q^\prime}) \, b^{\dagger}_\beta ({\bf q}) 
\  (2 \pi)^3 \delta ( {\bf p} + {\bf q}     - {\bf p^\prime} - {\bf q^\prime} )
\  \Big)~. 
\label{eq:H2primev2}
\end{eqnarray}

\end{widetext}

\bibliography{refs}

\end{document}